\newif\ifstandalone
\standalonetrue

\ifstandalone

\documentclass[11pt]{llncs}
\usepackage{graphicx}
\usepackage[round,sort]{natbib}

\let\oldendproof\endproof
\def\endproof{\qed\oldendproof}

\DeclareSymbolFont{AMSb}{U}{msb}{m}{n}
\DeclareSymbolFontAlphabet{\Bbb}{AMSb}
\def\Ree{\ensuremath{\Bbb R}}
\def\Zee{\ensuremath{\Bbb Z}}

\begin{document}
\title{Learning Sequences}
\author{David Eppstein}

\institute{Computer Science Department\\
Donald Bren School of Information \& Computer Sciences\\
University of California, Irvine\\
\email{eppstein@uci.edu}}

\maketitle

\begin{abstract}
We describe the algorithms used by the ALEKS computer learning system for manipulating combinatorial descriptions of human learners' states of knowledge, generating all states that are possible according to a description of a learning space in terms of a partial order, and using Bayesian statistics to determine the most likely state of a student. As we describe, a representation of a knowledge space using learning sequences (basic words of an antimatroid) allows more general learning spaces to be implemented with similar algorithmic complexity. We show how to define a learning space from a set of learning sequences, find a set of learning sequences that concisely represents a given learning space, generate all states of a learning space represented in this way, and integrate this state generation procedure into a knowledge assessment algorithm. We also describe some related theoretical results concerning projections of learning spaces, decomposition and dimension of learning spaces, and algebraic representation of learning spaces.
\end{abstract}

\else

\chapter{Learning Sequences}
\label{Learning Sequences}
\vspace*{-2cm}
{\large D. Eppstein\footnote{Dept. of Computer Science, University of California, Irvine}

\fi

\def\notdiv{\mathop{\not|}}

\section{Introduction}

ALEKS (short for Assessment and Learning in Knowledge Spaces) is a computer system, and a company built around that system, which helps students learn knowledge-based academic systems such as mathematics by assessing their knowledge and providing lessons in concepts that the assessment judges them as ready to learn.
\index{ALEKS}

Rather than being based on numerical test scores and letter grades, ALEKS is based on a combinatorial description of the concepts known to the student, in the form of a \emph{learning space} \citep{doign99}. In this formulation, feasible states of knowledge of a student are represented as sets of facts or concepts; the task of the assessment routine is to determine which facts the student knows. Not all sets of facts are assumed to form feasible knowledge states; therefore, if the assessment routine can find a sequence of questions the answers to each of which roughly halve the number of remaining states consistent with the answers, then the student's knowledge can be assessed from a number of questions approximately equal to the logarithm (base two) of the number of feasible knowledge states, a number that may be significantly smaller than the number of facts in the learning space. That is, informally, the learning space model allows the system to make inferences about the student's knowledge of concepts that have not been directly tested, from the information it has about the concepts that have been tested; these inferences can significantly reduce the number of questions needed to accurately assess the student's knowledge, and thereby greatly reduce the tedium of interacting with the system. In addition to speeding students' interactions with the system in this way, the combinatorial knowledge model used by ALEKS allows it to determine sets of concepts which it judges the student ready to learn, and present a selection of lessons based on those concepts to the student, rather than forcing all students to proceed through the curriculum in a rigid linear ordering of lessons.
\index{learning space}
\index{logarithm}
\index{Doignon, J.-P.}
\index{Falmagne, J.-Cl.}

The actual assessment and inference routine in the ALEKS system is based on a Bayesian formulation in which the system computes likelihoods of each feasible knowledge state from the students' answers, aggregates these values to infer likelihoods that the student can answer each not-yet-asked question, and uses these likelihoods to select the most informative question to ask next. Once the student's knowledge is assessed, the system generates a \emph{fringe} of concepts that it judges the student ready to learn, by calculating the feasible knowledge states that differ from the assessed state by the addition of a single concept, and presents the students with a selection of online lessons based on that fringe.
\index{Bayesian statistics}
\index{fringe!of a state}

As of its 2006 implementation, ALEKS's assessment procedure must run in Java, interactively on the user's PC, so efficient computation at interactive speeds is essential to its operation. As we describe in more detail in the next chapter, the ALEKS system uses a data representation for its learning spaces based on a partial order structure of prerequisites for each concept. A clever state generation procedure allows for the efficient generation of states in learning spaces of, typically, 50 to 100 facts. For larger learning spaces, the combinatorial explosion in the total number of states makes it infeasible to generate all states; instead, the system repeatedly samples a subset of the concepts in such a way that each unsampled concept is ``near'' a sampled one, generates a learning space from the restriction of the prerequisite partial order to the sampled concepts, assesses the student's knowledge of the sample, and uses that sampled assessment to refine the portion of the learning space within which further assessment is judged necessary.
\index{Java}
\index{partial order}
\index{state generation}
\index{prerequisite}

Although quite successful, this partial order based definition of a learning space suffers from some flaws. Primary among these flaws is a lack of adaptability: the structure of prerequisite orderings between concepts must be developed by human knowledge engineers, and is difficult to change in any automated way. From the pattern of responses to ALEKS's assessments, it may be possible to infer that some sets of facts are highly unlikely to be found as knowledge states of students; eliminating such states from the system would help reduce the number of questions required for assessment. More significantly, some concepts may not readily learnable by the students when the system judges that they are; these concepts should be removed from the selection of online lessons presented to the students to avoid the frustration of unlearnable lessons. Similarly, the pattern of student assessment answers may lead us to conclude that some additional states are present among the students but not available to the system's representation; adding these states to the system would allow for more accurate assessment procedures. The ability to modify the learning space represented in the system is of special interest in the context of internationalization of ALEKS, as we would expect the different educational systems and cultures in different countries to lead to students with different typical knowledge states. It would be of great interest to the ALEKS designers to develop automated adaption systems that can take taking advantage of ALEKS's large base of user data and reduce the human engineering effort needed to adapt the system to new concepts and new cultures. An automated adaption procedure, based on the generalization of the concept of a fringe from states in a learning space to learning spaces in a system of such spaces, was developed by \citet{Thi-01}; however, this adaption procedure was not used by ALEKS because the  partial order based learning space representation is insufficiently flexible to allow the generation of new learning spaces by the insertion and removal of states.
\index{adaptation}
\index{Thi{\'e}ry, N.}

A secondary flaw relates to the mathematical definition of learning spaces. The spaces representable by partial orders in the implementation of ALEKS do not comprise all possible learning spaces in the theory developed by \citet{doign99}, but rather form a significantly restricted class of spaces known as \emph{quasi-ordinal spaces}, in which the family of feasible sets can be shown to be closed under set unions and set intersections. Closure under set unions is justified pedagogically, both at a local scale of learnability (learning one concept does not preclude the possibility of learning a different one) and more globally (if student $x$ knows one set of concepts, and student $y$ knows another, then it is reasonable to assume that there may be a third student who combines the knowledge of these two students). However, closure under set intersections seems much harder to justify pedagogically: it implies that any concept has a single set of prerequisites, all of which must be learned prior to the concept. On the contrary, in practice it may be that some concepts may be learned via multiple pathways that cannot be condensed into a single set of prerequisites.  The practical effect of this flaw is that the partial order based representation for learning spaces used by ALEKS does not allow certain learning spaces to be defined; instead one must define a larger space formed by the family of all intersections of sets in the desired learning space, and the larger number of sets in this intersection closure of the learning space may lead to inefficiencies in the assessment procedure. In addition, and more seriously, the inability to accurately describe the prerequisite structure for certain concepts may lead to situations where the system incorrectly assesses a student as being ready to learn a concept.
\index{quasi-ordinal space}
\index{union!closure under}
\index{intersection!closure under}

In this chapter we outline algorithms and prototype implementations for a more flexible representation, one that would allow the implementation of Thi\'ery's automatic adaptation procedure and allow more general definitions of learning spaces than the existing partial order based representation allows, while preserving the scalability and efficient implementability of that representation. We believe that these goals may be achieved by using a representation based of \emph{learning sequences}: orderings of the learning space's concepts into linear sequences in which those concepts could be learned. A learning space may have an enormous number of possible learning sequences, but, as we show, it is possible to correctly and accurately represent any learning space using only a subset of these sequences, and in many cases the number of sequences needed to define the space can be very small. For instance, for the quasi-ordinal spaces currently in use by ALEKS, a representation based on learning sequences can be constructed in which the number of learning sequences equals the maximum number of concepts in the fringe of a feasible state. We show how to generate efficiently all states of a learning space defined from a set of learning sequences, allowing for similar and similarly efficient assessment procedures to the ones currently used by ALEKS. Additionally, we show how to find efficiently a representation of this type for any learning space, using an optimal number of example sequences, and how to adapt any space defined in this way by adding or removing sets from its family of feasible states. We detail this learning sequence based representation, and the efficient algorithms based on it, after describing in more detail ALEKS's existing partial order based representation.
\index{learning sequence}
\index{state generation}
\index{adaptation}

In addition we investigate more generally the theory of and algorithms for learning spaces and related combinatorial structures. In particular we examine the mathematical structure of projections of learning spaces, the extent to which it is possible to decompose learning spaces efficiently into unions of simpler learning spaces, definitions of learning spaces via the algebraic properties of their union operation, and relations between different definitions of dimension for a learning space. These theoretical investigations are detailed in later sections of this chapter.

\section{Learning Spaces from Partial Orders}

We outline in this section the representation of learning spaces already in use by the 2006 implementation of ALEKS.
As we describe, this representation leads to efficient assessment algorithms, but is only capable of representing a limited subset of the possible learning spaces, the so-called \emph{quasi-ordinal spaces}.
\index{quasi-ordinal space}

A \emph{partial order} is a relation $<$ among a set of objects, satisfying \emph{irreflexivity} ($x\not<x$) and
\emph{transitivity} ($x<y$ and $y<z$ implies $x<z$).
Although defined as a mathematical object, a partial order may be represented concisely for computational purposes by its \emph{Hasse diagram}, a directed graph containing an edge $x\rightarrow y$ whenever $x<y$ and there does not exist $z$ with $x<z<y$. That is, we connect a pair of items in the partial order by an edge whenever the pair belongs to the \emph{covering relation} of the partial order.  The original partial order may be recovered easily from the Hasse diagram representing it: $x<y$ if and only if there exists a directed path from $x$ to $y$ in the Hasse diagram.
\index{partial order}
\index{irreflexive relation}
\index{transitive relation}
\index{Hasse diagram}

To derive a learning space from a partial order on a set of concepts, we interpret the edges of the Hasse diagram as describing prerequisite relations between concepts. That is, if $x$ and $y$ are concepts, represented as vertices in a Hasse diagram containing the edge $x\rightarrow y$, then we take it as given that $y$ may not be learned unless $x$ has also already been learned. For instance, in elementary arithmetic, one cannot perform multi-digit addition without having already learned how to do single digit addition, so a learning space involving these two concepts should be represented by a Hasse diagram containing a path from the vertex representing single-digit addition to the vertex representing multi-digit addition.
\index{prerequisite}

With this interpretation, a state of knowledge in the learning space may be formed as a \emph{lower set}: a set $S$ of the concepts in a given partial order, satisfying the requirement that, for any edge $x\rightarrow y$ of the Hasse diagram, either $x\in S$ or $y\notin S$. Figure~\ref{fig:qos} shows an example of a Hasse diagram on eight concepts, and the 19 states in the learning space derived from this Hasse diagram.
\index{lower set}
\index{Hasse diagram}

\begin{figure}[t]
\centering\includegraphics[scale=0.45]{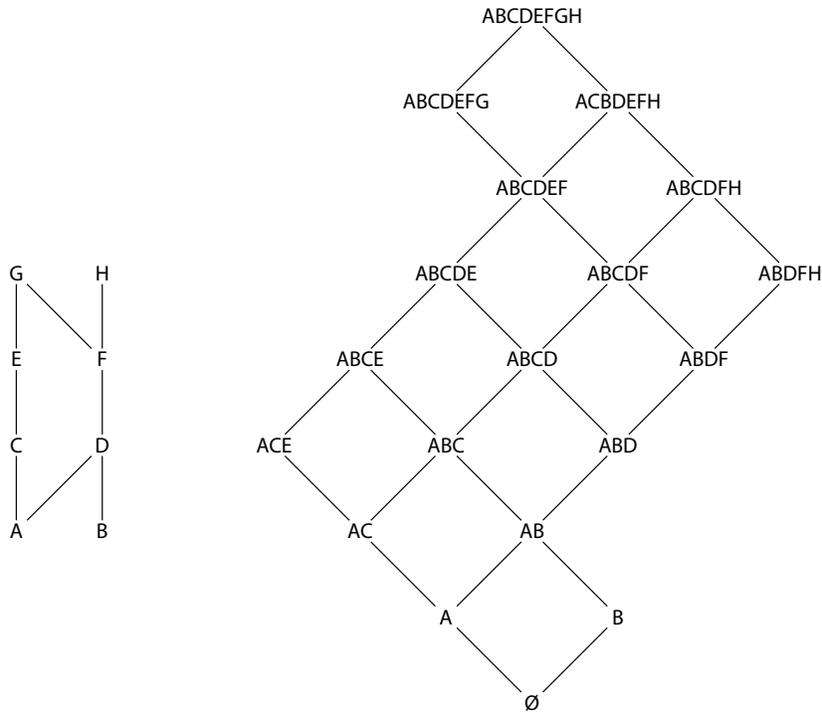}
\caption{Left: a partial order, shown as a Hasse diagram in which each edge is directed from the concept at its lower endpoint to the concept at its upper endpoint. Right: the learning space derived from the partial order on the left.}
\label{fig:qos}
\end{figure}

We call a learning space derived from a Hasse diagram in this way a \emph{quasi-ordinal space}. A quasi-ordinal space must satisfy the following three properties:
\index{quasi-ordinal space}

\begin{description}
\item[Accessibility.]
For every nonempty state $S$ in the learning space, there is a concept $x\in S$ such that $S\setminus\{x\}$ is also a state in the learning space. In learning terms, any state of knowledge may be reached by learning one concept at a time. ``Accessibility'' is the usual name for this property in the combinatorics literature;
in learning theory papers, it has also been referred to as ``downgradability'' \citep{doble01a}.
\index{accessible set family}
\index{downgradability}
\index{Doble, C.W.}
\index{Doignon, J.-P.}
\index{Falmagne, J.-Cl.}
\index{Fishburn, P.C.}

\item[Union Closure.]
If $S$ and $T$ are states of knowledge in the learning space, then $S\cup T$ is also a state in the learning space. In learning terms, the knowledge of two individuals may be pooled to form a state of knowledge that is also feasible.
\index{union!closure under}

\item[Intersection Closure.]
If $S$ and $T$ are states of knowledge in the learning space, then $S\cap T$ is also a state in the learning space. We are unaware of a natural learning based interpretation of this property.
\end{description}
\index{intersection!closure under}

A family of states satisfying only accessibility and union closure forms a mathematical structure known as an \emph{antimatroid} \citep{KorLovSch-91}, and it is this more general general class of structure that we hope to capture with our learning sequence representation of a learning space. The partial order based structure defined by ALEKS allows only a special subclass of antimatroids satisfying also the intersection closure property; mathematically such a structure forms a \emph{lattice of sets}, or, equivalently, a \emph{distributive lattice}.
Conversely, it follows from results of \citet{birkh37} that any distributive lattice can be represented via a partial order in this way. See \citet{doign99} for related representation theorems for learning spaces.
\index{antimatroid}
\index{lattice!of sets}
\index{lattice!distributive}
\index{Korte, B.}
\index{Lov{\'a}sz, L.}
\index{Schrader, R.}
\index{Doignon, J.-P.}
\index{Falmagne, J.-Cl.}
\index{Birkhoff, G.}
\index{Birkhoff's representation theorem}
\index{ALEKS}

In what follows we review briefly the algorithms used by ALEKS to perform assessments using this quasi-ordinal learning space representation.

\subsection{The Fringe}

The \emph{fringe} of any state $S$ in a learning space is defined to be the set of concepts that, when added to or removed from $S$, lead to another state in the learning space. Fringes are important to ALEKS because they describe the concepts that the assessed student is most ready to learn, or has the most shaky learning of. We may distinguish the \emph{outer fringe} of concepts a student is ready to learn, those concepts $x$ such that $S\cup\{x\}$ is also a state in the learning space, from the \emph{inner fringe} of concepts that a student may have most recently learned, those concepts $x$ such that $S\setminus\{x\}$ is also a state. The fringe is the union of the outer and inner fringes. In a learning space (or more generally, in any medium) each state may be uniquely identified by the pair of its outer and inner fringes.
\index{fringe!of a state}
\index{ALEKS}

In a quasi-ordinal space, the inner fringe of $S$ consists of the maximal concepts in $S$, and the outer fringe consists of the minimal concepts not in $S$. The fringe of $S$ may be calculated easily in a single pass through the edges $x\rightarrow y$ of the Hasse diagram: if $x\rightarrow y$ and $x\notin S$, then $y$ cannot be in the fringe, and if $x\rightarrow y$ and $y$ in $S$, then $x$ cannot be in the fringe.  We initialize a set $F$ to consist of the whole domain, and remove $x$ or $y$ from $F$ whenever we discover a covering relation that prevents it from belonging to the fringe; the remaining set at the end of this scan is the fringe.
\index{Hasse diagram}

\subsection{State Generation}

To determine the likelihood that a student knows each concept in a learning space, from a given set of results on asked questions, ALEKS uses an algorithm based on listing all states in the learning space. The algorithm used by ALEKS can be explained most easily in terms of \emph{reverse search} \citep{avis96}, a general technique for developing generation algorithms for many types of combinatorial structures.
\index{state generation}
\index{reverse search}
\index{ALEKS}
\index{Avis, D.}
\index{Fukuda, K.}

Suppose we have chosen a \emph{topological ordering} (also known as a \emph{linear extension}) of the concepts in a partial order; that is,
a sequence of the concepts such that, if $x<y$ in the order, then $x$ must appear prior to $y$ in the sequence. For instance, one may sort the concepts by the length of the longest path leading to each concept in the Hasse diagram, with ties broken arbitrarily among concepts having paths of the same length; the resulting sorted sequence is a topological ordering.
Then, given any state $S$ in the knowledge space, one may find another state $S\setminus\{x\}$ where $x$ is chosen to be the concept belonging to $S$ that has the latest position in the topological ordering. We call $S\setminus\{x\}$ the \emph{predecessor} of $S$. In this way, we disambiguate the accessibility property of learning spaces and make a concrete choice of which concept to remove to form the predecessor of any state.
If we repeat this removal process, starting from any state, we will eventually reach the empty set, so the graph formed by connecting each state to its predecessor is a tree, rooted at the empty set.
Reverse search, applied to this predecessor relationship, amounts to performing a depth first traversal of this tree.
\index{topological ordering}
\index{linear extension}
\index{partial order}
\index{predecessor}
\index{accessible set family}
\index{depth first search}
\index{Hasse diagram}

To be more specific, we perform a recursive traversal of the tree defined above, maintaining as we do for each state $S$ a set ${\rm children}(S)$ of the concepts that may be added to $S$ to form another state that is a child of $S$ in the predecessor tree. Then, when the recursion steps from $S$ to $S\cup\{x\}$, we calculate ${\rm children}(S\cup\{x\})$ from  ${\rm children}(S)$ by removing from ${\rm children}(S)$ any $y$ occurring prior to $x$ in the topological order, and adding to ${\rm children}(S)$ any concept $y$ reachable from $x$ by a Hasse diagram edge $x\rightarrow y$ such that all other prererequisites of $y$ already belong to $S$.
Once we have calculated ${\rm children}(S\cup\{x\})$, we may output $S\cup\{x\}$ as one of our states and continue recursively to each state $S\cup\{x,y\}$ for each $y$ in ${\rm children}(S\cup\{x\})$, and so on.
\index{children of a state}
\index{Hasse diagram}

Very little is needed in the way of data structures beyond the Hasse diagram itself to implement this recursive traversal efficiently. Primarily, we need a way of quickly determining whether all prerequisites of some concept $y$ belong to the set $S$ we are traversing.  This may be done efficiently by maintaining, for each $y$, a count of the prerequisites of $y$ that do not yet belong to $S$, decrementing that count for each successor of $x$ whenever we step from $S$ to $S\cup\{x\}$, and incrementing the count again when our recursive traversal returns from $S\cup\{x\}$ to $S$. In this way we may test prerequisites in constant time, and use time proportional to the number of Hasse diagram edges out of $x$ to update counts whenever we step from state to state. Alternatively, the method used within the 2006 implementation of ALEKS is to store a bitmap representation of $S$, and mask it against a bitmap representing the predecessors of $y$ whenever such tests are needed; theoretically this method requires time proportional to the number of items in the learning space per test, but in practice it is fast because typical modern computer architectures allow for the testing of prerequisite relations for 32 concepts simultaneously in a single machine instruction.
\index{prerequisite}
\index{predecessor}
\index{bitmap}
\index{ALEKS}

\begin{figure}[t]
\centering\includegraphics[width=4in]{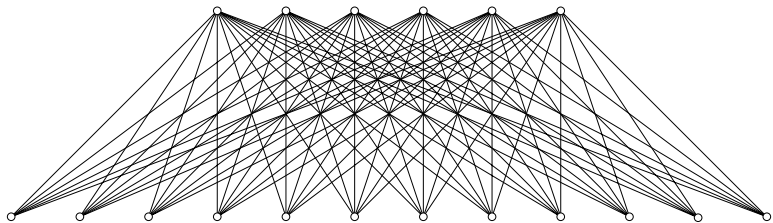}
\caption{The Hasse diagram of a worst case example for the partial order state generation algorithm.
Each of $2n/3$ items (bottom) is connected by an edge to each of $n/3$ items (top).
The corresponding learning space has $2^{2n/3}+2^{n/3}-1$ states, $2^{2n/3}$ of which cause the state generation algorithm to perform $n/3$ prerequisite checks per state, so the total time is $\Omega(n)$ per state.}
\label{fig:slowqos}
\end{figure}

The total time spent copying lists of children in this procedure can be charged using amortized time analysis against the time to generate each child state. Therefore, the bottleneck for time analysis of the procedure is the part of the time spent testing successors of $x$ and testing whether to add those successors to ${\rm children}(S\cup\{x\})$. As described above, each such test can be implemented in linear time, so the total time for the algorithm is proportional to the sum, over all states in the learning space, of the number of Hasse diagram edges that are outgoing from the last concept in each state. In typical examples the Hasse diagrams are relatively sparse and the time per state may approach a constant, but even in the worst case (Figure~\ref{fig:slowqos}) the time is no more than $O(n)$ per state in a learning space with $n$ concepts. It is this level of state generation efficiency that we hope to approach or meet with our more general learning space representation.
\index{amortized time analysis}
\index{worst case time complexity}
\index{state generation}
\index{children of a state}
\index{Hasse diagram}

\subsection{Assessment Procedure}

While a student is being assessed, he or she will have answered some of the assessment questions correctly and some incorrectly. We desire to infer from these results likelihoods that the student understands each of the concepts in the knowledge space, even those concepts not explicitly tested in the assessment.

The inference method used by ALEKS \citep{falmagne:88a,falma90} applies more generally to any system of sets, and can be interpreted using Bayesian probability methods.
We begin with a prior probability distribution on the feasible states of the learning space. In the most simple case, we can assume an uninformative uniform prior in which each state is equally likely, but the method allows us to incorporate prior probabilities based on any easily computable function of the state, such as the number of concepts in the set represented by the state. These prior probilities may incorporate knowledge about the student's age or grades, or results from assessments in previous sessions that the student may have had with ALEKS; for instance, if a previous session assessed the student as being in a certain state, we could use an a priori probability distribution based on distance from that state in our next assessment. However, in the 2006 implementation of ALEKS only uniform prior probabilities are used.
\index{Bayesian statistics}
\index{prior probability distribution}
\index{uniform prior}
\index{ALEKS}
\index{session}
\index{Falmagne, J.-Cl.}
\index{Doignon, J-P.}
\index{Koppen, M.}
\index{Villano, M.}
\index{Johanessen, L..}

We also assume a conditional probability that the student, in a given state, will answer a question correctly or incorrectly: if the question tests a concept within the set represented by the state, we assume a high probability of a correct answer, while if a question tests a concept not within the set represented by the state, we assume a low probability of a correct answers.
Answers opposite to what the given state predicts can be ascribed to careless mistakes or lucky guesses, and we assume that incidences of such answers are independent of each other. ALEKS' test questions are designed so that lucky guesses are very rare; therefore, the necessity for accurate assessment in the presence of careless mistakes is of much greater significance to the design of the assessment procedure, but this also means that it is necessary to ascribe different rates to these two types of events. With these assumptions, we may use Bayes' rule to calculate posterior probabilities of each state. To do so, we calculate for each state a likelihood, the product of its prior probability with a set of terms, one term per already-asked assessment question. The value of each term depends on whether the student answered the question correctly and whether the concept tested by the question belongs to the given state. Once we have calculated these likelihoods for all states, they may be converted to probabilities by dividing them all by a common constant of proportionality, the sum of the likelihoods of all states.
\index{careless mistake}
\index{lucky guess}
\index{Bayes' rule}
\index{constant of proportionality}
\index{partition function}

From these posterior probabilities of states we wish to calculate posterior probabilities of individual concepts. To do so, we sum the probabilities of all states containing a given concept. ALEKS's assessment procedure calculates the probabilities of all concepts in this way, and then chooses to test the student on the most informative concept: the one with probability closest to 50\% of being known. Eventually, all concepts will have probabilities bounded well away from 50\%, at which point the evaluation procedure terminates.
\index{posterior probability}

Although described above as a separate sum for each concept,
ALEKS implements this probability calculation via a single pass through the states of the learning space. When the state generation procedure steps from a state $S$ to a child state $S\cup\{x\}$, it calculates the likelihood (product of terms for each answered question) from the new state by multiplying the old likelihood by a single term for the questions based on concept $x$. It totals the likelihood of all states descending from $S\cup\{x\}$ in the recursive search, and adds this total likelihood to that of concept $x$. Then, when returning from $S\cup\{x\}$ to $S$, it adds the total likelihood calculated for $S\cup\{x\}$ into that for $S$. In this way, the likelihood for each concept is calculated in constant time per state, and the constant of proportionality needed to turn these likelihoods into probabilities is calculated as the total likelihood at the root of the recursion.

If $x<y$ in the partial order defining the system's learning space, $x$ will be assessed as having higher probability than $y$ of being known; therefore, the set of concepts returned by this assessment algorithm is guaranteed to be a feasible knowledge state for the given quasi-ordinal space.

\subsection{Hierarchical Sampling Scheme}

Although the assessment procedure described above works well for partial orders of 50 to 100 concepts, it becomes too slow for larger learning spaces due to the need for the algorithm to list all states in the space and the combinatorial explosion in the number of states generated for those spaces. Therefore, the ALEKS system resorts to a sampling scheme that allows its assessment procedure to run at interactive speeds for much larger learning spaces. This sampling scheme is based on three concepts, all depending on the details of the definition of the learning space in terms of partial orders: distance between concepts, definition of smaller learning spaces from sampled concepts, and bounding concept likelihoods from their prerequisites and postrequisites.
\index{combinatorial explosion}
\index{ALEKS}

To generate smaller samples of the set of concepts used to define a learning space, ALEKS uses a notion of distance between two concepts in a partial order. To define the distance between $x$ and $y$,
define $\Delta_{x,y}$ to be the set of items whose comparison to $x$ is different from its comparison to $y$. That is,
\begin{eqnarray*}
\Delta_{x,y}=\{x,y\}&\cup&
\{z\mid z<x \wedge z\not<y\}\quad \cup \quad
\{z\mid z<y \wedge z\not<x\}\\
&\cup&\{z\mid x<z \wedge y\not<z\}\quad \cup\quad
\{z\mid y<z \wedge x\not<z\}.
\end{eqnarray*}
Then the distance $d(x,y)$ between $x$ and $y$ is defined to be $|\Delta_{x,y}|-1$.
This distance satisfies the mathematical axioms defining a \emph{metric space}: $d(x,x)=0$, $d(x,y)=d(y,x)$. For any $x$, $y$, and $z$,
$\Delta_{x,z}\subset\Delta_{x,y}\cup\Delta_{y,z}$, and the union is not disjoint as $y$ belongs to both sides, so $d(x,z)\le d(x,y)+d(y,z)$.
\index{distance!between concepts}
\index{metric space}

ALEKS then chooses a suitable distance threshhold $\delta$, and a sample $S$ of the concepts of the learning space such that every concept is within distance $\delta$ of a member of $S$. Although there is no mathematical proof of such a fact, the intent of this sampling technique is that assessment on a nearby sample concept is likely to be informative about the assessment of each unsampled concept.

\begin{figure}[t]
\centering\includegraphics[scale=0.45]{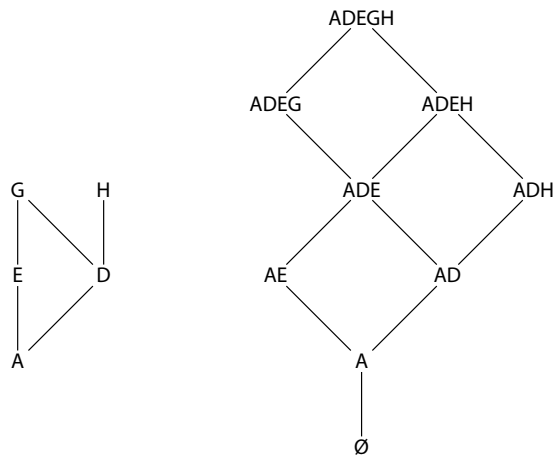}
\caption{The Hasse diagram of the partial order of Figure~\ref{fig:qos}, restricted to the sampled set of concepts $\{A,D,E,G,H\}$, and the smaller learning space generated from the restricted partial order.}
\label{fig:sampleqos}
\end{figure}

Once a sample of concepts has been chosen, ALEKS must form a learning space describing the knowledge states possible for that sample, so that it can apply its assessment procedure to the sample.
For quasi-ordinal learning spaces, this process of forming a learning space on the sample is very simple: one need merely restrict the given partial order defining the space to the sampled concepts, and build a learning space from the restricted order (Figure~\ref{fig:sampleqos}).
\index{restriction!of a partial order}

Finally, the assessment of likelihoods on the sampled concepts is used to bound the likelihoods of the remaining unsampled concepts, to determine which ones the student is likely to know or not to know.
If $x<y$, $y$ belongs to the sample, and the student knows $y$ with probability $p$, then the student is taken to know the easier concept $x$ with probability at least $p$. Similarly, if $x<y$, $x$ belongs to the sample, and the student knows $x$ with probability $p$, then the student is taken to know the harder concept $y$ with probability at most $p$. However, there is something of a mismatch between these likelihood bounds and the distance-based sampling procedure: it is possible for the nearby samples to an unsampled concept $x$ to all be incomparable to $x$, in which case we cannot find any useful bounds for the likelihood of $x$.

This sampling process, sample learning space construction, and likelihood bound, are used together repeatedly to refine the portion of the learning space that is relevant for the student. Initially, all states are considered relevant, and a sample with a high distance threshhold is chosen. After several steps of refinement, a larger number of concepts have likelihoods that can be bounded away to one side or another of 50\%, and a sample is chosen with a smaller distance threshhold among only those remaining informative concepts. Eventually, this refinement process converges with all concepts having likelihoods bounded away from 50\%, which we may use to construct a most likely knowledge state for the student.

For learning spaces not defined from partial orders, we may have to replace these constructions with alternative techniques. However, it will still be necessary to have some way of sampling the concepts of the learning space, building a smaller learning space from the sample, and using assessments on the sample to bound likelihoods of unsampled concepts, because this sampling procedure is crucial to limiting the number of states generated in the assessment procedure and thereby keeping the assessment calculation's times fast enough for human interaction.

\section{Learning Spaces from Learning Sequences}

We now describe an alternative method for defining and describing learning spaces, that we believe may form the basis for an efficient and more flexible implementation of learning space based knowledge assessment algorithms than the one currently used by ALEKS.

While there has been past work on algorithmic characterizations of learning spaces using the terminology of antimatroids \citep{BoyFai-DAM-90,KemLev-arXiv-03}, that work focuses on showing that certain algorithms work correctly if and only if the structure they are applied to is an antimatroid. Here instead our focus is on implementation details for allowing antimatroid-based algorithms to run efficiently.
\index{antimatroid}
\index{Boyd, E.A.}
\index{Faigle, U.}
\index{Kempner, Y.}
\index{Levit, V.E.}

\subsection{Learning Sequences}

Given any learning space, there may be many orderings through which a student, starting from no knowledge, could learn all the concepts in the space. We call such an ordering a \emph{learning sequence}; in the combinatorics literature these are also known as \emph{basic words}.  Formally, a learning sequence $\sigma$ can be defined as a one-to-one map from the integers $\{0,1,2,\ldots n-1\}$ to the $n$ concepts forming the domain of a learning space, with the property that each \emph{prefix} $P_i(\sigma)=\sigma(\{0,1,2,\ldots i-1\})$ is a valid knowledge state in the learning space.  The sequence of prefixes $P_0(\sigma),P_1(\sigma),\ldots P_{n-1}(\sigma)$ forms a shortest path in the learning space from the empty set to the whole domain, and for any such path the sequence of items by which one state in the path differs from the next forms a learning sequence.
\index{learning sequence}
\index{basic word}
\index{prefix}

For instance, in the learning space depicted in Figure~\ref{fig:qos}, the leftmost path from the bottom state (the empty set) to the top set (the whole domain) passes through the sequence of states
$\emptyset$, $\{A\}$, $\{A,C\}$, $\{A,C,E\}$, $\{A,B,C,E\}$, $\{A,B,C,D,E\}$, $\{A,B,C,D,E,F\}$, $\{A,B,C,D,E,F,G\}$, and $\{A,B,C,D,E,F$, $G,H\}$. The learning sequence corresponding to this path is $A,C,E$, $B,D,F,G,H$.
Similarly, the learning sequence corresponding to the rightmost path in the figure is $B,A,D,F,H,C,E,G$.
Altogether, the learning space in Figure~\ref{fig:qos} can be shown to have 41 distinct learning sequences.

\begin{figure}[t]
\centering\includegraphics[scale=0.45]{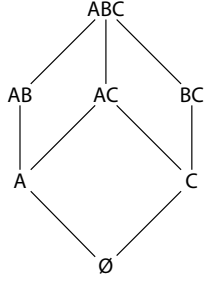}
\caption{A learning space that is not defined from a partial order. Unlike partial order based learning spaces, this space is not closed under intersections: $\{A,B\}$ and $\{B,C\}$ are both states in the space, but their intersection $\{B\}$ is not. However, the space still satisfies the union closure and accessibility requirements of learning spaces.}
\label{fig:nob}
\end{figure}

For a quasi-ordinal learning space, such as the one in Figure~\ref{fig:qos}, a learning sequence is essentially just a topological ordering of the partial order on concepts defining the quasi-ordinal space. However, we have defined learning sequences in a way that can be applied to any learning space. For instance, the learning space in Figure~\ref{fig:nob}, which does not come from a partial order, has four learning sequences: (1) $A,B,C$, (2) $A,C,B$, (3) $C,A,B$, and (4) $C,B,A$.
\index{quasi-ordinal space}
\index{topological ordering}

\subsection{States from Sequences}

If we are given a set $\Sigma$ of learning sequences, we can immediately infer that all prefixes of $\Sigma$ are states in the associated learning space. But we can also infer the existence of other states using the union-closure property of learning spaces: any set formed from a union of prefixes of $S$ must be a state in the associated learning space.
\index{union!closure under}

For instance, suppose we have the two sequences $A,B,C$ and $C,B,A$ from the learning space in Figure~\ref{fig:nob}. The prefixes of these sequences are the six sets $\emptyset$, $\{A\}$, $\{A,B\}$, $\{A,B,C\}$, $\{C\}$, and $\{B,C\}$. However, by forming unions of prefixes we may also form the seventh set $\{A\}\cup\{C\}=\{A,C\}$. All seven states of the learning space can be recovered in this way from unions of the prefixes of these two sequences.

In general, for any set of sequences $\Sigma$ over a domain of concepts, define the learning space ${\cal L}_\Sigma$ as the family of unions of prefixes of $\Sigma$.
If $\Sigma$ consists of learning sequences from a learning space $\cal L$, then
${\cal L}_\Sigma\subset\cal L$. We will discuss, in a later section of this chapter, methods of selecting a small set $\Sigma$
such that ${\cal L}_\Sigma=\cal L$. For now, we take $\Sigma$ as given and describe the learning space ${\cal L}_\Sigma$ it generates.

\begin{figure}[t]
\centering\includegraphics[width=4.5in]{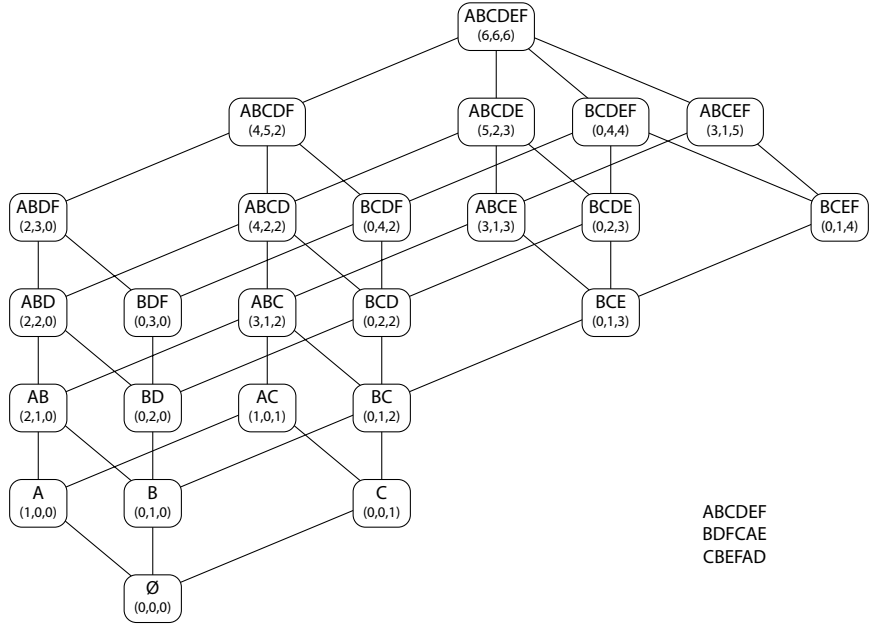}
\caption{Learning space ${\cal L}_\Sigma$ generated from three sequences
$ABCDEF$, $BDFCAE$, and $CBEFAD$. Each state is shown with its index ${\rm mex}(S)$.}
\label{fig:seqex}
\end{figure}

\subsection{Indexing States}

It is possible to name states in ${\cal L}_\Sigma$ by vectors in $\Zee^{|\Sigma|}$, in such a way that each state has a unique name and the concepts in each state can be reconstructed easily from the state's name. Such a naming scheme will be useful in several of our later algorithms.
\index{vector}

Given a state $S$ in a learning space, and a set
$$\Sigma=\{\sigma_0,\sigma_1,\ldots,\sigma_{k-1}\}$$
of $k$ learning sequences within the space, define ${\rm mex}_i(S)$ to be the minimum index in $\sigma_i$ of a concept excluded from $S$; that is,
$${\rm mex}_i(S)=\min\{j\mid \sigma_i(j)\notin S\}.$$
If $S$ is the whole domain, we define for completeness ${\rm mex}_i(S)=n$. 
Equivalently, therefore, ${\rm mex}_i(S)$ is the size of the largest prefix of $\sigma_i$ that is a subset of $S$.
Define ${\rm mex}(S)$ to be the vector
$${\rm mex}(S)=({\rm mex}_0(S),{\rm mex}_1(S),\ldots,{\rm mex}_{k-1}(S)).$$
That is, {\rm mex} can be viewed as a function mapping states in the learning space to vectors in $\Zee^k$.
\index{minimum excluded element}

Conversely, we can define a function ${\rm up}(v)$, mapping vectors in $\Zee^k$ to states in the learning space, by
$${\rm up}(v)=\bigcup_{0\le i<k} P_{v_i}(\sigma_i).$$
That is, we interpret the coordinates of $v$ as lengths of prefixes of each of the sequences in $\Sigma$, and form the union of these prefixes.  For any set $S$,
${\rm up}({\rm mex}(S))\subset S$, with ${\rm up}({\rm mex}(S))=S$ if and only if $S$ is a state in the learning space ${\cal L}_\Sigma$.

If ${\rm mex}(S)$ is known, ${\rm mex}(S\setminus\{x\})$ may easily be calculated, by 
taking the coordinatewise minimum of ${\rm mex}(S)$ and the positions of $x$.
However, calculating ${\rm mex}(S\cup\{x\})$ from ${\rm mex}(S)$ appears to be more complicated.
As part of our state generation procedure, described in a later section, we include a bitmap-based data structure that allows us to maintain a set subject to both insertions and deletions and calculate its ${\rm mex}$ indices, more quickly than recalculating these indices from scratch each time they are needed.
\index{state generation}
\index{bitmap}

The indices ${\rm mex}(S)$ for an example of a learning space ${\cal L}_\Sigma$ are depicted in Figure~\ref{fig:seqex}. The learning space shown in the figure cannot be derived from a partial order, as it has states $\{B,D,F\}$ and $\{B,C,E,F\}$ but not their intersection $\{B,F\}$.

\subsection{Axioms of Learning Spaces}

As we now argue, the space ${\cal L}_\Sigma$ defined from a set $\Sigma$ of learning sequences automatically satisfies the union-closure and accessibility properties that we require of our learning spaces.

Union-closure follows from the indexing scheme we have defined above:
if $S$ and $T$ are states in ${\cal L}_\Sigma$, then
$$S\cup T={\rm up}(\max({\rm mex}(S),{\rm mex}(T))),$$
where in this formula $\max$ should be interpreted as the pointwise maximum of two vectors.
Thus, $S\cup T$ is also a state in ${\cal L}_\Sigma$.
\index{union!closure under}

\begin{figure}[t]
\centering\includegraphics[width=4in]{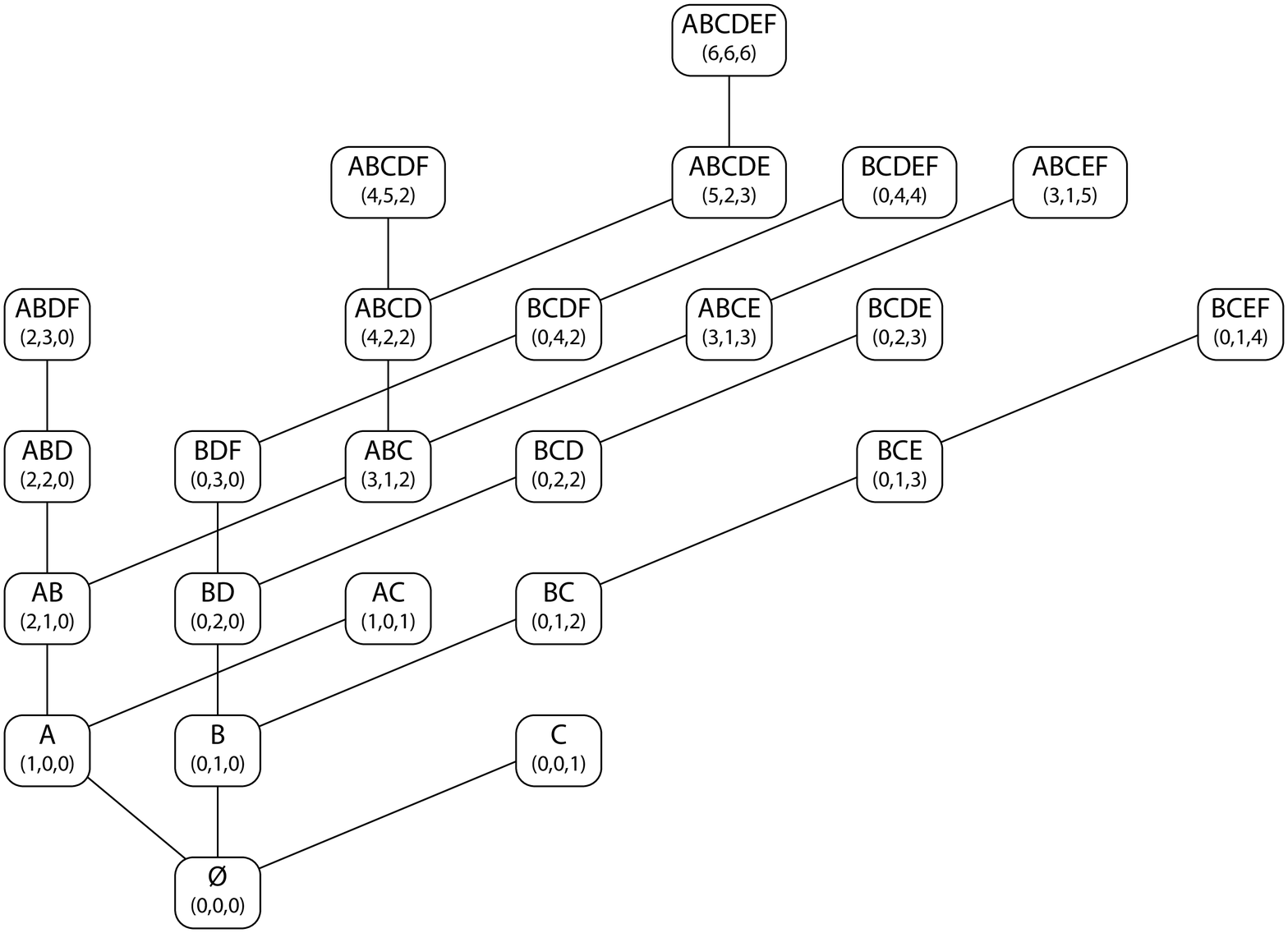}
\caption{Predecessors of the states in the learning space from Figure~\ref{fig:seqex}.}
\label{fig:seqpred}
\end{figure}

Accessibility also follows from the indexing scheme.
For any nonzero vector $v\in\Zee^k$, let ${\rm decrement}(v)$ be the vector formed by decreasing the last nonzero coordinate of $v$ by one, and let ${\rm decrement}^i(v)$ be the result of iterating the decrement operation $i$ times. Let
$${\rm diff}(v)=\min\{i\mid {\rm decrement}^i(v)\ne v\},$$ 
and, for any nonempty state $S$ in ${\cal L}_S$
let
$${\rm pred}(S) =
{\rm up}({\rm decrement}^{{\rm diff}({\rm mex}(v))}({\rm mex}(v)).$$
Figure~\ref{fig:seqpred} shows a link from $S$ to ${\rm pred}(S)$ for each state $S$ in the learning space depicted in Figure~\ref{fig:seqex}.
Then, each iteration of ${\rm decrement}$ causes the image of the vector under the ${\rm up}$ operation either to stay unchanged or to change by the removal of a single item. Since ${\rm pred}(S)$ is defined to be the image after the first iteration that causes a change, it differs from $S$ by the removal of a single item, so the requirement of accessibility is achieved.
\index{accessible set family}

Conversely, any learning space satisfying union closure and accessibility can be defined as ${\cal L}_\Sigma$ for an appropriate set $\Sigma$ of learning sequences. In a later section we describe how to choose $\Sigma$ to be as small as possible, as the size of $\Sigma$ will be directly related to the efficiency of the algorithms we use to assess learning in ${\cal L}_\Sigma$.

\subsection{The Fringe}

As with learning spaces derived from partial orders, we will need to calculate fringes of states.
The \emph{outer fringe} of $S$, the items that may be added to state $S$ to form a new state,
is particularly easy to describe: Each state $S\cup\{x\}$ may be formed as the union of the prefixes defining $S$ itself, together
with one more prefix $P_{{\rm mex}_i(S)}(\sigma_i)$, so the outer fringe of $S$ is
$$\{\sigma_i({\rm mex}_i(S))\mid 0\le i<k\}.$$
The \emph{inner fringe} of $S$, those items in $S$ that may be removed to form another state,
is a little more complicated, but still easily expressible using our indexing scheme: it is
$$\{x\in S\mid {\rm up}({\rm mex}(S\setminus\{x\}))=S\setminus\{x\}\}.$$
The formula ${\rm up}({\rm mex}(S\setminus\{x\}))=S\setminus\{x\}$ is true if and only if $S\setminus\{x\}$ is a state of the learning space, so we may construct the lower fringe by testing whether this formula is true for each member $x$ of~$S$.
\index{fringe!of a state}

\section{Efficient State Generation and Assessment}

We wish to generalize ALEKS's knowledge assessment algorithms to learning spaces derived from learning sequences, so that we may use these spaces in place of the quasi-ordinal spaces used in ALEKS. The keys to this generalization are methods for efficiently listing all states of a learning space, and for projecting learning spaces onto smaller subsets of concepts in order to speed up the algorithm even further by reducing the size of the state space. As we describe here, a representation based on learning sequences allows us to perform these operations efficiently.

\subsection{State Generation}

As with partial orders, we can list all states of a sequence-based learning space by an algorithm based on exploring the tree generated by the predecessor operation we defined earlier; refer back to Figure~\ref{fig:seqpred} for a depiction of this tree on an example learning space.
Recall that the predecessor of a state is generated by repeatedly reducing the last nonzero coordinate of its ${\rm mex}$ vector until the ${\rm up}$ function maps the reduced vector to a set different from the state itself. In order to reverse this, and find the successors of a state $S$ (that is, the states having $S$ as its predecessor), we need merely perform the following steps:
\index{predecessor}
\index{state generation}
\index{reverse search}
\index{minimum excluded element}

\begin{enumerate}
\item Set $p$ to the smallest value of $i$ such that $S$ is the union of prefixes from the first $i$ sequences.
\item For each $i\ge p$, such that $\sigma_i({\rm mex}_i(S))\ne \sigma_j({\rm mex}_j(S))$ for all $j<i$
(that is, such that the first excluded concept in $\sigma_i$ differs from the first excluded
concept in all earlier sequences):
\begin{enumerate}
\item Let $v$ be formed from ${\rm mex}(S)$ by adding one to its $i$th coordinate.
\item Let $T={\rm up}(v)$ and output $T$.
\item Recursively generate states starting from $T$.
\end{enumerate}
\end{enumerate}

\noindent
It is not hard to show that every state generated by this procedure has $S$ as its predecessor, and that all states having $S$ as predecessor will be generated by this procedure.
\index{predecessor}

To generate all states in the learning space, we perform a depth first traversal of the states of the space, by applying this procedure recursively starting with the empty set.
If $T$ is a successor of $S$, the value of $p$ in the search for the successors of $T$ equals the value of $i$ used when we generated $T$ from its parent $S$; therefore, we may pass these $p$ values through the recursive traversal and avoid calculating them at each step.  However, the ${\rm mex}$ values of $S$ and $T$ may differ not just in the $i$th coordinate but also in other coordinates greater than $i$, and must be recalculated.
If the learning space is generated by $k$ learning sequences, the time per state is $O(k)$,
except for the time to calculate the ${\rm mex}$ value of each new state.
\index{depth first search}
\index{worst case time complexity}

\subsection{Bitmap Structure for Minimal Excluded Elements}

In order to implement our state generation procedure efficiently, we need to be able to calculate the ${\rm mex}$ of each newly generated state. When we defined the ${\rm mex}$ operation we showed that removing an element from a state allows the new ${\rm mex}$ to be calculated easily, but, unfortunately, in the case of the state generation procedure, each new state is generated not by removing an element but by adding one.
\index{minimum excluded element}

To speed up this part of the state generation, we maintain as part of the recursive traversal a collection of long integers, one per learning sequence defining the learning space. In each integer $B_i$ we store the number $\sum\{2^j\mid\sigma_j(i)\in S\}$ representing the set of positions of learning sequence $\sigma_j$ that are present in the current state $S$. Each of these integers may be updated in constant time for each step forwards and backwards in the state traversal, and each ${\rm mex}$ calculation may be performed by finding the first zero bit in the binary representation of each integer,
which may be performed efficiently using a combination of arithmetic and bitwise boolean operations on the integer. Specifically, the bitwise exclusive or of $B_i$ and $B_i+1$ is an integer of the form $2^j-1$, where $j-1$ is the position of the first zero in $B_i$ and should be used as the value of ${\rm mex}_i(S)$.
\index{bitmap}

In a programming language without built-in support for integers of arbitrarily large precision, it may be appropriate to replace each value $B_i$ in this structure by an array of 32-bit machine integers, each one representing the intersection of $S$ with some 32-member subset of the concepts.

\subsection{Bayesian Assessment}

Given the state generation procedure, we can assess the likelihoods of a student knowing each concept using the same Bayesian procedure described for quasi-ordinal spaces.
\index{Bayesian statistics}
\index{prior probability distribution}

Specifically, we calculate a likelihood for each state of the space as a product of the prior probability of the state with a set of terms, one term per question asked of the student. As we generate all states, we can calculate each such likelihood as the product of the likelihood of a previously generated state by a single term. The likelihood of a concept is the sum of the likelihoods of the states containing it, and the probability of knowing a concept is this sum normalized by dividing by the sum of likelihoods of all states.

As for the assessment procedure in quasi-ordinal spaces, our state generation procedure operates recursively by adding single concepts to previously generated states. Therefore, if we maintain the sum of likelihoods in each subtree of the recursion, and add each such sums to the total likelihood of the most recently added item, we may compute the sums of likelihoods for all concepts with a single pass over all states of the learning space, in constant additional time per step.

\subsection{Smaller Spaces on Subsets of Concepts}

\begin{figure}[t]
\centering\includegraphics[width=3.5in]{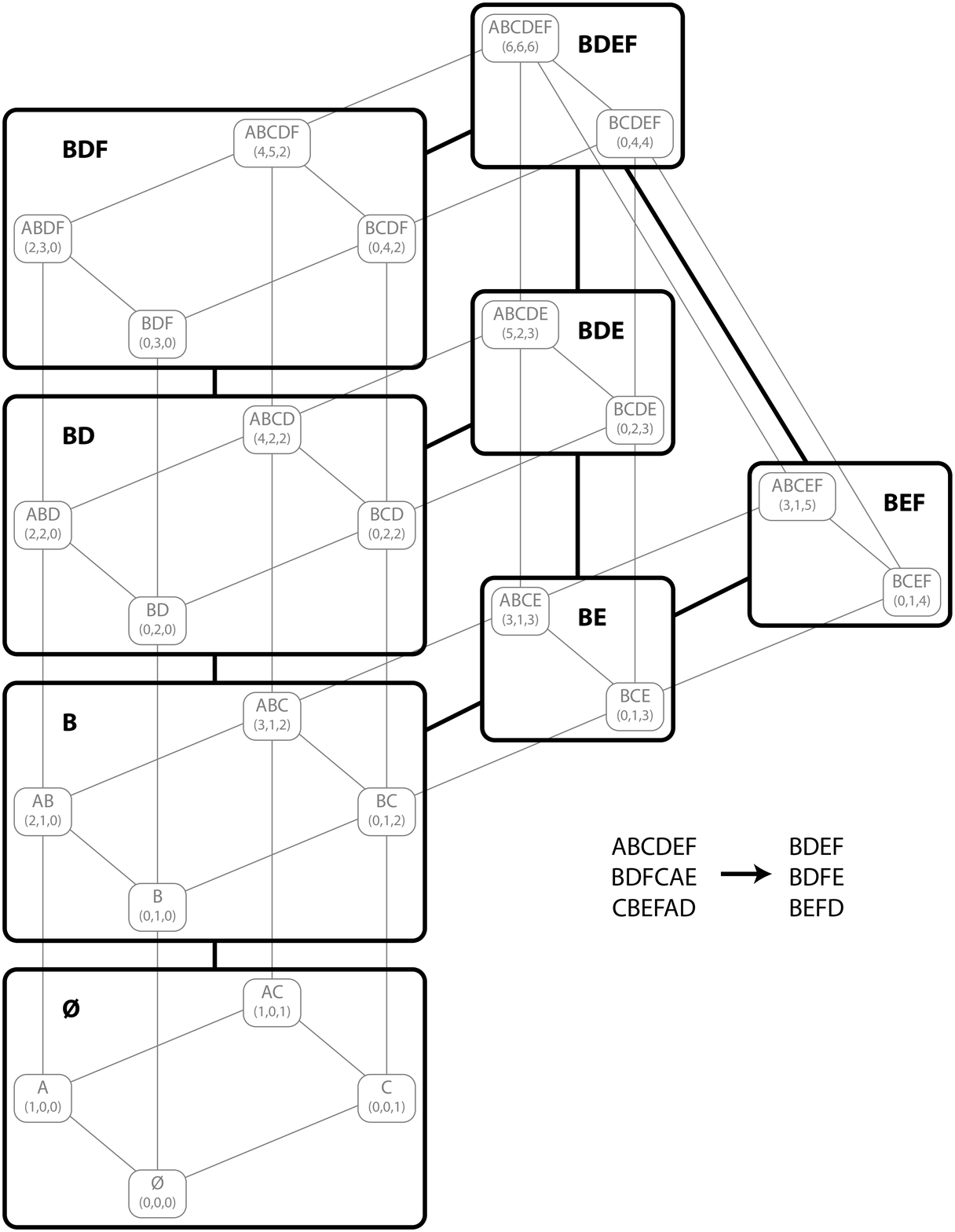}
\caption{The projection of a learning space.}
\label{fig:seqproj}
\end{figure}

If $\cal F$ is any family of sets, and $P$ is any subset of $\cup\cal F$, we may define
the \emph{projection}
$${\cal F}_P=\{S\cap P\mid S\in{\cal F}\}$$
by intersecting each state in ${\cal F}$ with $P$.
If $\cal F$ is a learning space, then so is any projection of it. The projection inherits the union closure property of learning spaces from $\cal F$,
as $(S_1\cap P)\cup(S_2\cap P)=(S_1\cup S_2)\cap P$. To prove the accessibility property of learning spaces
for a set $S\cap P$, $S\in\cal F$, repeatedly use accessibility in $\cal F$ to remove concepts from $S$ until a concept in $P$ is removed, and let the predecessor of $S\cap P$ be formed by removing the same concept from $S\cap P$.
\index{projection!of a learning space}
\index{union!closure under}
\index{accessible set family}
\index{predecessor}

If $\cal L$ is a quasi-ordinal space, the projection ${\cal L}_P$ is also quasi-ordinal order, formed from the restriction of the partial order defining $\cal L$ to the items in $P$. This concept of a learning space derived from a restricted partial order (without the more general notion of projection) is used by ALEKS, as described earlier, to speed up its assessment procedures. Similarly, if $\cal L$ is a learning space derived from a collection of learning sequences, the projection  ${\cal L}_P$ can be derived from the subsequences formed by listing the items of $P$ in the order they appear in each sequence defining $\cal L$, as shown in Figure~\ref{fig:seqproj}. Therefore, we may efficiently construct a learning sequence based representation of a projection of a learning space defined by learning sequences. This projected space may not necessarily use the minimum possible number of sequences in its representation, but it uses a number of sequences no more than that of the larger space it was projected from.
\index{quasi-ordinal space}

\subsection{Projection for Faster Assessment}

The choice of how many concepts to use in modeling knowledge as a learning space is somewhat arbitrary: we may model some areas of knowledge coarsely, as a relatively small set of concepts, and others more finely using many concepts. We do not expect our assessment of the likelihood that a student knows a given concept to vary significantly based on the choice of how coarsely the space is modeled. Therefore, though there is little mathematical justification for this simplification, we may greatly speed up our knowledge assessment routines by projecting to a space consisting only of the relevant concepts for the assessment: those we wish to assess, and those for which we have answers.

To be specific, suppose that $Q$ is a set of concepts on which we have tested a student's knowledge, in a learning space $\cal L$, and $x$ is a concept for which we wish to infer a likelihood of the student's knowledge.
Rather than applying the state-generation based assessment algorithm to the whole learning space $\cal L$, we may apply it to the projected space ${\cal L}_{Q\cup\{x\}}$.
\index{projection!of a learning space}

Mathematically, in order to get assessment results in the projection equal to the results in the original space $\cal L$, we would need to use a prior probability in each projected state $S$ equal to the sum of the prior probabilities of the states in $\cal L$ that project to $S$. For instance, if we are using the uninformative prior for $\cal L$, in which we assume that each state is equally likely to occur, then in the projection we should use a prior in which the probability for each state $S$ is proportional to the number of states in $\cal L$ that project to $S$. However, even for the uninformative prior, this problem of quickly calculating numbers of states seems quite difficult. Instead, we propose performing the evaluation in the projected space  ${\cal L}_{Q\cup\{x\}}$ as if it were the whole space. The results of the likelihood calculation in this projection may differ from those of the calculation in the whole space $\cal L$, but there seems little reason in principal to view one of the results as more reliable than the other, and if $Q$ is small then the projected assessment may be much more efficient than the assessment in the original space. If $Q$ is so small that ${\cal L}_Q$ does not provide adequate coverage of the learning space, however,
the assessment may be less informative than that in $\cal L$; for instance, concepts occurring in very few states of the learning space would have very low likelihood when assessed in $\cal L$ with $Q=\emptyset$ but $50\%$ likelihood when assessed in ${\cal L}_{\{x\}}$. This problem of a too-small sample losing information about the number of states containing concept $x$ may be ameliorated by adding a small random sample $R$ of the learning space concepts to $Q$; that is, if we perform the assessment in ${\cal L}_{Q\cup R\cup\{x\}}$ in place of ${\cal L}_{Q\cup\{x\}}$, we can gain an accurate estimate of the number of states containing $x$ even when $Q$ is empty or near-empty.

One unfortunate feature of computing assessments in ${\cal L}_{Q\cup\{x\}}$ rather than $\cal L$ is that it requires a separate run of the state generation procedure for each separate concept~$x$, while the assessment procedure in ${\cal L}$ involves only a single run of the state generation procedure. It seems likely that, if we wish to compute likelihoods for all concepts in the knowledge space,
a faster procedure would be to partition the concepts into small collections $C_i$, and to use the state generation based assessment procedure on ${\cal L}_{Q\cup C_i}$ to compute likelihoods for each concept in $C_i$. If $C_i$ is chosen to be too small, this would involve many runs of the procedure, while if $C_i$ is chosen to be too large, then the procedure will list too many states, so there is likely some optimal intermediate value (perhaps depending on the size of $Q$) to use for the size of the collections $C_i$ in order to make this projection based assessment procedure run at optimal speed. The tradeoffs of runtime versus collection size would need to be found by experimental analysis beyond the scope of this chapter. However, if the collections $C_i$ are chosen randomly rather than as distance based clusters, they would also perform the same function as the set $R$ above of making
the states of the projected space more representative of the numbers of states in the original space. Thus, it may be desirable for the accuracy of the assessment to chose $C_i$ larger than an analysis based only on algorithmic efficiency would suggest.

We are hopeful that the speedup provided by this projection based assessment method, relative to the more naive state generation based assessment used by the current version of ALEKS, would allow student assessment to proceed in a single pass involving all concepts in the learning space, rather than the multipass clustering based approach used in the current version of ALEKS. In outline, the procedure would proceed as follows:

\begin{enumerate}
\item Initialize $Q$ to $\emptyset$.
\item Repeat:
\begin{enumerate}
\item Partition the concepts of the learning space randomly into collections $C_i$.
\item For each $C_i$, generate the states of ${\cal L}_{Q\cup C_i}$ and apply the assessment algorithm to compute likelihoods for each concept in $C_i$.
\item If all concepts have likelihoods bounded away from $50\%$, terminate the assessment procedure.
\item Choose a concept $x$ with likelihood as close as possible to $50\%$, assess the student's knowledge of $x$, and add $x$ to $Q$.
\end{enumerate}
\end{enumerate}

As discussed above, the choice of how large to make the collections $C_i$ is left undetermined by this algorithm, and involves a tradeoff between algorithmic efficiency and assessment accuracy.

\subsection{Inverting Projections to Shrink the State Space}

We considered also an alternative assessment procedure to the one described in the previous section, in which we maintain sets of concepts $K$ and $U$ which we believe the student to know or not know respectively, and restrict the state space search used in the assessment algorithm to only those states consistent with $K$ and $U$. That is, rather than totalling likelihoods for all states in ${\cal L}$, we only
total those likelihoods in the states of ${\cal L}$ that project to $K$ in ${\cal L}_{K\cup U}$; all other states are assumed to have a priori probability zero of being the actual state of knowledge of the student. As $K$ and $U$ grow through the course of the assessment, this could lead to a significant reduction in the number of states that need to be listed.  However, some form of sampling and assessment based on the projection of $\cal L$ onto the sample would still be needed in order to achieve adequate performance in the earlier stages of the algorithm when $K$ and $U$ are small. One possible structure for such an algorithm is shown below.

\begin{enumerate}
\item Initialize $K$ and $U$ to empty.
\item Repeat:
\begin{enumerate}
\item Choose a sample $S$ of the concepts in $\cup{\cal L}\setminus(K\cup U)$.
\item Repeat:
\begin{enumerate}
\item Use the likelihood calculation algorithm on ${\cal L}_{S\cup K\cup U}$, restricted to
only those states that project to $K$ in ${\cal L}_{K\cup U}$, to compute likelihoods for $S$.
\item If no concept in $S$ has likelihood near $50\%$, terminate the inner loop.
\item Choose a concept with likelihood as close to $50\%$ as possible and test the student's knowledge of that concept.
\end{enumerate}
\item Add the concepts of $S$ with likelihoods significantly larger than $50\%$ to $K$, and add the concepts of $S$ with likelihoods significantly smaller than $50\%$ to $U$.
\item If all concepts of ${\cal L}$ are in $K$ or $U$, terminate the algorithm.
\end{enumerate}
\end{enumerate}

This algorithm appears to improve on the previous section's algorithm by requiring only a single
pass through the states of a projected learning space per question asked of the student. However we see several significant potential problems with this approach that stopped us from investigating it in greater detail. First, the assignment of states to $K$ and $U$ is made with certainty, making a Bayesian interpretation of this algorithm problematic. Second, while it is possible to generate the states consistent with $K$ and $U$ in relatively small time per state, we have not been able to find an algorithm for this task that is as efficient as the one for listing all states in a learning space. And third and most troubling, if the algorithm ever reaches a situation where the projection ${\cal L}_{K\cup U}$ does not contain $K$ as a state, the algorithm will fail by dividing by zero in the calculation of the concept likelihoods, so it is not robust against inaccuracies in the learning space model or the assessment algorithm.

Nevertheless if we wish to implement such an algorithm we must find the set of states in $\cal L$ project to $K$ in ${\cal L}_{S\cup K\cup U}$, in order to perform the likelihood calculations of the algorithm.
We may impose a tree structure on this set of states by choosing a learning sequence $\sigma$ of $\cal L$ that contains as a prefix the union of the states that project to $K$, and defining the parent of any state $S$ to be the state formed by adding to $S$ the first missing concept in $\sigma$. A reverse search algorithm that reverses this parent relation and traverses this tree can list all states that project to $K$, in a small number of steps per listed state, but we have not found a way to make this algorithm as efficient as the one we gave earlier for listing all states in a learning space. We omit the details.

\subsection{Distances and Clustering}

We have provided above a method based on projection that does not involve the distance-based clustering currently used in ALEKS. However, it may still be of interest to define a distance function on the concepts of the learning space. For instance, as we have discussed above, clustering based on distance may be useful for the partition into collections $C_i$ in our projection based assessment algorithm. In addition, the distance from some previously assessed state may be an important ingredient in the prior probability distribution on learning space states used by the knowledge assessment algorithm.
\index{distance!between concepts}

The learning sequences we use to define our learning spaces provide a natural family of distances:
if we interpret each concept's position in each sequence as a coordinate, all such positions form a vector corresponding to the concept in $\Zee^k$, and we may use any $L_p$ metric on this collection of vectors. We did not investigate more carefully the choice of metric to determine the value of $p$ that would work best for knowledge assessment applications.

\section{Finding Concise Representations}

We have seen that a learning space may be defined from a set $\Sigma$ of learning sequences.
The running time of our algorithms depends on the number of sequences in $\Sigma$, so in order for this representation of a learning space to yield efficient algorithms, we need this number to be small.

For instance, for the learning space in Figure~\ref{fig:nob}, we have seen that two of the four learning sequences, $A,B,C$ and $C,B,A$, suffice to determine all seven states in the learning space.
For the learning space in Figure~\ref{fig:qos}, two learning sequences, 
$A,C,E,B,D,F,G,H$ and $B,A,D,F,H,C,E,G$, corresponding to the leftmost and rightmost paths through the diagram in the figure, similarly determine the whole learning space; the remaining 39 sequences are superfluous. More generally, we have shown \citep{eppst06} that two learning sequences suffice to describe a learning space if, and only if, the space can be drawn as a planar graph with the empty set and the whole domain as vertices on the outer face of the drawing. In that paper we outlined a complicated algorithm for finding such a pair of sequences, when it exists, in time linear in the number of states of the learning space.
\index{Eppstein, D.}
\index{graph drawing}

We show here that we may efficiently find the minimum possible set $\Sigma$ defining any given learning space $\cal L$. This minimum number is known in antimatroid theory as the \emph{convex dimension} of an antimatroid, and it can be defined as the \emph{width} of a certain partial order associated with the learning space \citep{KorLovSch-91}.  Our technique applies known polynomial time algorithms for calculating widths of partial orders.
\index{convex dimension}
\index{width!of a partial order}
\index{Korte, B.}
\index{Lov{\'a}sz, L.}
\index{Schrader, R.}

For learning spaces defined from partial orders themselves, the convex dimension is the width of the original partial order from which the learning space was defined. This implies that, even for these special learning spaces, our algorithms are similarly efficient to the Hasse-diagram based algorithms currently in use by ALEKS.

\subsection{The Width of a Partial Order}

A \emph{chain} in a partial order is a set of items that are all comparable to each other; any chain can be ordered into a sequence $x_0,x_1,\ldots$ such that $x_i<x_j$ if and only if $i<j$. A \emph{chain cover} is a set of chains that together include all items in the order.
Similarly, an \emph{antichain} in a partial order is a set of items no two of which are comparable to each other, so that the partial order places no restrictions on their ordering.
The \emph{width} of a partial order is the maximum cardinality of any of its antichains.
It is well known (Dilworth's Theorem) that, for any partial order, the width is also equal to the minimum number of chains in a chain cover.
\index{antichain}
\index{chain cover}
\index{width!of a partial order}
\index{Dilworth's theorem}
\index{Dilworth, R.P.}

For instance, the partial order depicted on the left of Figure~\ref{fig:qos} can be covered by two chains, $\{A,C,E,G\}$ and $\{B,D,F,H\}$. It also has many antichains of two concepts, for instance $\{A,B\}$. Therefore, its width is exactly two. There cannot be any antichain of three or more concepts, nor is there a single chain that covers all its concepts.

\begin{figure}[t]
\centering\includegraphics[width=4in]{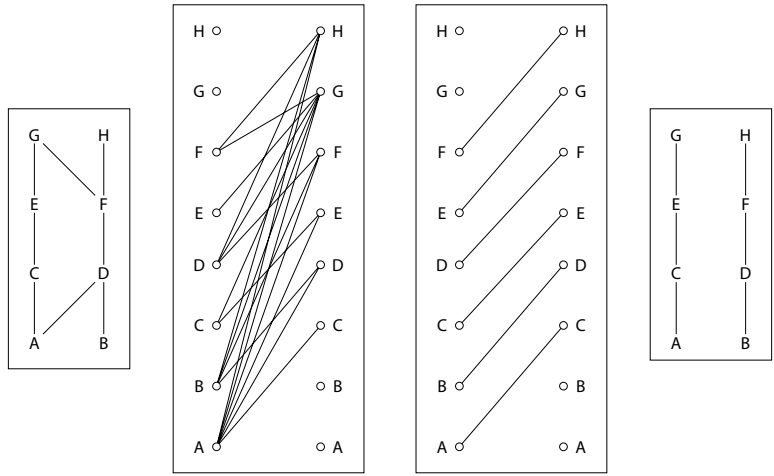}
\caption{An optimal chain decomposition of a partial order may be found via bipartite graph matching.}
\label{fig:bip4cd}
\end{figure}

An optimal chain decomposition of a given partial order may be found by a technique based on bipartite graph matching (Figure~\ref{fig:bip4cd}).
The far left of the figure shows the same partial order as the one in Figure~\ref{fig:qos}, and the center left shows a bipartite graph derived from this order. In the bipartite graph, we have two vertices for each item in the partial order, one on the left and one on the right. We draw an edge from the vertex labeled $x$ on the left to the vertex labeled $y$ on the right whenever $x<y$. In general this graph may be disconnected, and have vertices incident to no edges; these complications cause no difficulty for our algorithms.
A \emph{matching} in a graph is a set of edges, at most one edge incident on any vertex, and a \emph{maximum matching} is a matching maximizing the total number of edges in the set.
A maximum matching of the graph in the figure is shown in center right.
From any matching in this graph, we may derive a chain cover, by including in the same chain any two items in the partial order that form the two endpoints of a matched edge; such a cover, for the matching in the figure, is shown on the far right. In a partial order with $n$ items, matchings with $k$ matched edges correspond to chain covers with $n-k$ chains, so the minimum chain cover can be found from the maximum matching.
\index{maximum matching}
\index{bipartite graph}

\citet{HopKar-SJC-73} showed how to find a maximum matching in a bipartite graph with $n$ vertices and $m$ edges in time $O(mn^{1/2})$; for partial orders with $n$ items this translates to an $O(n^{5/2})$ time bound for finding an optimal chain decomposition and therefore finding the width. A maximum antichain may also be found in the same amount of time; we omit the details as they are not necessary for our learning space application.
\index{chain cover}
\index{antichain}
\index{worst case time complexity}
\index{Hopcroft, J.E.}
\index{Karp, R.M.}

\subsection{Learning Sequences for Chains of States}

Recall that a \emph{chain} is a totally ordered collection of objects in a partial order.
Therefore, a chain of states of a learning space consists of a nested family of sets.
We show here, for $\cal C$ any chain of states in learning space $\cal L$,
how to find a learning sequence $\sigma$ in $\cal L$ such that each state of $\cal C$
is a prefix of $\sigma$. We may assume without loss of generality that $C$ contains $\emptyset$ and $\cup\cal L$.
\index{learning sequence}

To find $\sigma$, sort the sets in $\cal C$ by size:
$$\emptyset=S_0\subset S_1\subset S_2\subset\cdots\subset S_k=\cup\cal L.$$
We will build up $\sigma$ in stages; after the $i$th stage,
we will have chosen values for $\sigma_0,\sigma_1,\ldots\sigma_{|S_i|-1}$ such
that all sets $S_j$, $j\le i$, are prefixes of the chosen sequence of values
and such that all prefixes of the sequence are states in $\cal L$.
Initially, the empty sequence satisfies this requirement for $i=0$.

In stage $i$, we will have already generated a sequence for the elements of $S_{i-1}$.
By the accessibility property of learning spaces, we can also generate a sequence
$\tau_0,\tau_1,\ldots\tau_{|S_i|-1}$ of the elements of $S_i$, such that each prefix of this sequence is a state in $\cal L$. We then form our new longer sequence of $\sigma_j$ values by concatenating 
the subsequence of the $\tau_j$'s not already chosen as $\sigma_j$'s onto the end of the previously chosen $\sigma_i$'s. This concatenation preserves the property that the sets $S_j$, $j<i$,
are all prefixes of the new sequence, and now $S_i$ itself is the prefix consisting of the whole concatenation. In addition, each prefix of the new sequence is the union of a prefix of the old sequence of $\sigma_j$'s and of a prefix of the sequence of $\tau_j$'s, and therefore by the union closure property of learning spaces is a state of $\cal L$.

Once we have completed the final stage of this process, for $S_k=\cup\cal L$,
the ordering on $\sigma$ provides the desired learning sequence.
We refer to the existence of this learning sequence $\sigma$ as the chain property of learning spaces.

In general, the algorithm described here involves a number of predecessor calculations equal to the sum of the cardinalities of the sets in the chain. 
For learning spaces derived from partial orders on $n$ concepts, described by Hasse diagrams with $m$ edges, a faster algorithm is possible, based on a standard method for performing topological ordering via depth first search: we sort the elements in descending order by the size of the smallest chain set they belong to, and  perform a depth first traversal of the Hasse diagram, initiating the traversal at the elements in the sorted order. The desired learning sequence is the reverse of a postorder numbering for this traversal. Thus, after ordering the elements by the smallest containing chain sets, the remaining algorithm takes time $O(m+n)$.
\index{worst case time complexity}
\index{Hasse diagram}
\index{predecessor}

\subsection{The Base of a Learning Space}

The \emph{base} of a union-closed family of sets is a minimal subfamily such that any member of the family can be reconstructed as a union of members of the subfamily.
For \emph{antimatroids} (union-closed accessible families) there is a particularly easy construction of a base.
\index{antimatroid}
\index{base}

Define a \emph{predecessor} of set $S$ in antimatroid $\cal A$ to be any set $T=S\setminus\{x\}$ that also belongs to $\cal A$. If $S$ has zero predecessors, and $\cal A$ is accessible, then $S$ must be empty and can be reconstructed as an empty union of sets no matter what base is chosen.
And, if $S$ has two or more predecessors, each of which can be reconstructed as a union of sets in the base, then $S$ itself can be reconstructed as the union of any two of its predecessors.
Thus in either of these two cases $S$ cannot belong to a base.
However, if $S$ has only a single predecessor $T$, then any proper subset of $S$ is also a subset of $T$: by the chain property, there must be a learning sequence containing the proper subset and $S$ itself as prefixes, and the predecessor of $S$ on that sequence must be $T$. Therefore, in this case, any union of subsets of $S$ forms a subset of $T$ and cannot equal $S$, so $S$ must belong to every base.
From these considerations we see that any antimatroid $\cal A$ has a unique base ${\cal B}({\cal A})$ consisting of the sets that have a single predecessor. We view ${\cal B}({\cal A})$ as forming a partial order by set inclusion: if $S$ and $T$ are both sets in the base, then $S<T$ in the partial order if and only if $S\subset T$.
\index{predecessor}

\begin{figure}[t]
\centering\includegraphics[scale=0.45]{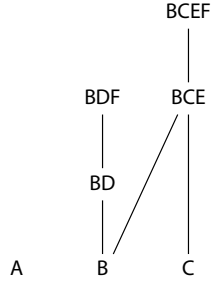}
\caption{The Hasse diagram of the base ${\cal B}({\cal L}_\Sigma)$ for the learning space ${\cal L}_\Sigma$ from Figure~\ref{fig:seqex}.}
\label{fig:seqbasis}
\end{figure}

In a learning space $\cal L$ derived from a partial order, the base sets have the form $S_x=\{y\mid y<x\}\cup\{x\}$. For, the only predecessor of $S_x$ is $S_x\setminus\{x\}$, but any set not of this form is either empty or has two or more independent maximal elements, each of which can be removed to form a predecessor. Also, $S_x\subset S_y$ if and only if $x<y$. Thus the base sets correspond one-for-one with the concepts of $\cal L$ and the partial order on ${\cal B}({\cal L})$ is isomorphic to the partial order on concepts. However, for learning spaces not derived from partial orders, ${\cal B}({\cal L})$ may contain many more sets than there are concepts in the learning space.
Figure~\ref{fig:seqbasis} depicts the partial order formed by the base of the learning space depicted in Figure~\ref{fig:seqex}; this base has seven sets although it describes a learning space on only six concepts.
\index{predecessor}

It will also be of interest to find the base of a learning space ${\cal L}_\Sigma$ defined from a set $\Sigma$ of learning sequences. For learning spaces of this type, the only states that could be in a base are the prefixes of $\Sigma$: any state that is not a prefix can be formed as a union of smaller sets, and therefore cannot be in the base.
If state $S$ is a prefix of one of our defining sequences $\sigma_i$, with $x$ as the latest concept in that prefix, then $S$ is a base if and only if the position of $x$ in each other $\sigma_j$ is either the same as its position in $\sigma_i$ or later than the minimum excluded element of $S$ in $\sigma_j$.
For, if this is so, no other symbol can be removed from $S$ leaving a valid state, while if it is not true
then there exists a symbol $y$ such that ${\rm mex}(S\setminus\{y\})$ is not dominated by ${\rm mex}(S\setminus\{x\})$, and the $y$ for which  ${\rm mex}(S\setminus\{y\})$ is maximal among other such vectors may be removed to form another valid state.
Thus, we may test whether $S$ is a base state, once its ${\rm mex}$ has been calculated,
in time proportional to the number of defining sequences.
We may find the base by listing the prefixes of each defining sequence, in order from longer prefixes to smaller ones so that we may more efficiently update the ${\rm mex}$ values, and applying this test to each prefix. In a space with $n$ concepts, defined by $k$ sequences, this base generation algorithm takes time $O(nk^2)$ and generates at most $kn$ base sets.
\index{base}
\index{prefix}
\index{worst case time complexity}

\subsection{Which Learning Sequences Define the Space?}

If $\Sigma$ is a collection of learning sequences for a learning space $\cal L$,
then ${\cal L}_\Sigma\subset{\cal L}$.  When does the learning space ${\cal L}_\Sigma$ derived from $\Sigma$ equal ${\cal L}$?

If every base set for $\cal L$ is a prefix of $\Sigma$, then all states in $\cal L$ can be formed as unions of those base members, and ${\cal L}_\Sigma={\cal L}$.  Conversely, if some set $B\in{\cal B}({\cal L})$ is not a prefix of $\Sigma$,
then $B$ can not be reconstructed as a union of any other sets in $\cal L$, let alone of the prefixes of $\Sigma$, and ${\cal L}_\Sigma\ne{\cal L}$.
\index{base}
\index{prefix}

The prefixes of a single learning sequence form a chain in the set inclusion ordering, and the base elements within that family form a chain in ${\cal B}({\cal L})$. Conversely, if we have any chain in ${\cal B}({\cal L})$, we may use the chain property to form from it a learning sequence containing as prefixes all chain members.  Thus, a set $\Sigma$ of learning sequences determines the learning space ${\cal L}_\Sigma={\cal L}$ if and only the base prefixes for each chain form a chain cover of ${\cal B}({\cal L})$.
The minimum number of learning sequences needed to determine the learning space is the minimum number of chains in a chain cover of ${\cal B}({\cal L})$, that is, its width.
\index{chain cover}
\index{width!of a partial order}

\subsection{Calculating the Convex Dimension}

We are now ready to describe an algorithm that, given a learning space $\cal L$, calculates its convex dimension and finds a minimum family $\Sigma$ of learning sequences such that ${\cal L}_\Sigma=\cal L$. The algorithm simply performs the following steps:
\index{convex dimension}

\begin{enumerate}
\item Find the base ${\cal B}({\cal L})$.
\item Use bipartite matching to find a minimum chain cover of ${\cal B}({\cal L})$.
\item Use the chain property to extend each chain in the cover to a learning sequence for $\cal L$.
\item Let $\Sigma$ be the set of learning sequences so constructed. The convex dimension is the number of learning sequences in~$\Sigma$.
\end{enumerate}

If only the convex dimension and not the learning sequences themselves is desired, step~3 can be omitted.
Assuming a suitably algorithmic implementation of the underlying learning space $\cal L$, all of these steps may be performed in time polynomial in $|{\cal B}|$ and in the number of concepts in $\cal L$,
both of which we expect to be far smaller than the total number of states in $\cal L$.

For learning spaces derived from partial orders on $n$ concepts, described by Hasse diagrams with $m$ edges, the first step is trivial, and the second step takes time $O(n^{5/2})$. There $k\le n$ chains in the cover,
and any single chain may be extended to a learning sequence in time $O(m)$. Thus, the third step can be implemented in time $O(mk)$, and the overall algorithm takes time $O(n^{5/2}+mk)$.
index{worst case time complexity}
\index{Hasse diagram}

Once a minimum family $\Sigma$ of learning sequences for $\cal L$ has been generated, we may redefine $\cal L$ as ${\cal L}_\Sigma$, and use our computational representation based on learning sequences for any future calculations in $\cal L$.

\section{Adapting a Learning Space}

\citet{Thi-01} defines the \emph{fringe} of a learning space to be the collection of sets that can be added to the space as new states, or removed from the states of the space, in order to form new learning spaces with more or fewer states. He calculates the fringes of a space as part of an algorithm for adapting a knowledge space to make it more accurately reflect the observed knowledge of students. One of the great advantages of learning sequence based construction of learning spaces, over partial orders, is their ability to allow such adaption: learning sequences can represent any learning space, and in particular can represent the spaces formed by adding states to and removing states from an existing space. Such adaptivity seems a very useful capability to add to ALEKS, as a knowledge space that more accurately models student knowledge would improve its interactions with students, and as ALEKS has in its user base a readily available supply of information about the accuracy of its model. We feel, however, that for learning spaces with numbers of states as large as those used by ALEKS, it would take impractically many moves to adapt a learning space to another one sufficiently different as to make a significant difference to ALEKS' assessment algorithms. Therefore, there seems plenty of scope for investigation of adaptivity algorithms that change more than one state of a space at a time. Nevertheless we describe here how to calculate the fringes of a learning space defined from learning sequences.
\index{adaptation}
\index{fringe!of a learning space}
\index{ALEKS}
\index{Thi{\'e}ry, N.}

As with the fringe of an individual state in a learning space, we may distinguish two subsets of the fringe of the learning space itself.  The \emph{inner fringe} of $\cal L$ consists of those states of $\cal L$ that may be removed, leaving a learning space with fewer states, while the \emph{outer fringe} of $\cal L$ consists of those sets that are not states of $\cal L$ but may be added as states, resulting in a learning space with more states.

\subsection{The Inner Fringe}

State $S$ may be removed from learning space $\cal L$ if, and only if, it satisfies all of the following requirements:
\begin{enumerate}
\item $S$ is in the base of $\cal L$.
\item $S\ne\cup\cal L$.
\item No set $S\cup\{x\}$, $x\notin S$, is in the base of $\cal L$.
\end{enumerate}
If $S$ were not in the base, it could be formed as the union of other sets in $\cal L$, and removing it would violate union closure. Similarly, if some set $S\cup\{x\}$ were in the base, removing $S$ would violate accessibility for that set. On the other hand, if the requirements above are all satisfied, then the space formed by removing $S$ from $\cal L$ satisfies union closure and accessibility, and therefore forms a valid learning space. The base of this new space consists of the remaining base members of $\cal L$, together with certain sets of the form $S\cup\{x\}$, from which we may use our concise representation algorithm to construct a representation of ${\cal L}\setminus\{S\}$ in terms of a small number of learning sequences.  All removable sets may be found efficiently, for spaces constructed from learning sequences, by listing all base sets and testing the above conditions.
\index{base}

\subsection{The Outer Fringe}

The generation of sets that may be added to a learning space $\cal L$ to form an augmented learning space is somewhat more problematic, as it cannot be done efficiently by generating and modifying the base of $\cal L$. If $S$ is any state of $\cal L$, the sets $S\cup\{x\}$ that can be added to $\cal L$ are exactly those for which $x$ belongs to the intersection of the outer fringes of states of the form $S\cup\{y\}$.
For, such an $x$ cannot belong to a state $S\cup\{x\}$ already, as the outer fringe of that state would not contain $x$; adding $S\cup\{x\}$ as a state preserves accessibility through $S$; and membership in the intersection ensures the union closure property of the augmented space. On the other hand, if $x$ does not belong to the outer fringe of some $S\cup\{y\}$, then $S\cup\{x\}$ cannot be added to $\cal L$, as the resulting space would not contain the union of $S\cup\{x\}$ and $S\cup\{y\}$.

Thus, to generate all sets that may be added to $\cal L$, we need merely list all states of $\cal L$, list for each state $S$ the states $S\cup\{y\}$ by computing the outer fringe of $S$, and intersect the outer fringes of the states $S\cup\{y\}$. Each such set is listed only once by this procedure, as the set $S$ for which it is listed must be unique or it would already belong to $\cal L$ by union closure. If we add a set $S\cup\{x\}$ as a new state to a learning space $\cal L$, the base of the new learning space consists of the newly added set, together with all base members of $\cal L$ that are not of the form $S\cup\{x,y\}$.
\index{state generation}

\section{Theoretical Investigations}

We describe here the results of some investigations on the mathematics of and algorithms for learning spaces, less directly related to efficient and flexible knowledge assessment. We state our results here in a more formal theorem-proof style than we use for the rest of this chapter.
 
\subsection{Fibers of Projections}

We discussed briefly earlier the possibility of forming reduced state spaces by inverting projections: if we believe that a student knows the concepts in a set $K$, and that the student doesn't know the concepts in a set $U$, what structure can we see in the subset of learning space states ${\cal L}(K,U)$ consistent with those beliefs? The states in ${\cal L}(K,U)$ are exactly those that project to $K$ in ${\cal L}_{K\cup U}$. Therefore, we can view ${\cal L}(K,U)$ as a \emph{fiber} of the projection, the inverse image of $K$. What combinatorial properties does such a fiber have?
\index{projection}
\index{fiber}

We observe that, more generally, if $\cal F$ is any family of sets, and ${\cal F}(K,U)$ denotes the subfamily of sets in $\cal F$ that are supersets of $K$ and disjoint from $U$, then closure under unions or intersections of $\cal F$ leads to the same property of ${\cal F}(K,U)$. For, taking the union or intersection of two sets in ${\cal F}(K,U)$ cannot form a set that is not a superset of $K$ or not disjoint from $U$.
Similarly, if $\cal F$ is well-graded, then ${\cal F}(K,U)$ inherits the same property.
Learning spaces derived from partial orders can be characterized as well-graded set families that are closed under unions and intersections, so if $\cal L$ is a learning space formed in this way, so is
${\cal L}(K,U)$.

For general learning spaces, however, a fiber need not itself form a learning space. For instance, for the learning space $\cal L$ of Figure~\ref{fig:nob}, the projection ${\cal L}_{\{B\}}$ has two states, $\emptyset$ and $\{B\}$, but the inverse image ${\cal L}({B},\emptyset)$ of $\{B\}$ is the family of three states $\{\{A,B\},\{A,B,C\},\{B,C\}\}$, which does not form a learning space as it fails the accessibility property. However, the fiber of a learning space projection is well-graded and closed under unions, and therefore forms a \emph{closed medium} \citep{falma02}.
\index{medium!closed}
\index{Falmagne, J.-Cl.}
\index{Ovchinnikov, S.}

If $\cal L$ is a learning space, we define an \emph{upper subfamily} ${\cal L}^+$ of $\cal L$ to be a subset of the states of $L$, such that if $S\subset T$ are two states of $\cal L$ with $S\in{\cal L}^+$ then $T\in{\cal L}^+$.
\index{upper subfamily}

\begin{theorem}
Any fiber ${\cal L}(K,U)$ forms an upper subfamily of a learning space ${\cal L}'$,
and any upper subfamily ${\cal L}^+$ of a learning space $\cal L$ can be represented as a fiber ${\cal L}'(K,U)$ for some learning space ${\cal L}'$ and sets $K$ and $U$.
\end{theorem}

\begin{proof}
First, let ${\cal L}(K,U)$ be a fiber, and let $S=\cup{\cal L}(K,U)$. Then $S$ is the unique maximal
state of ${\cal L}(K,U)$, and ${\cal L}(K,U)={\cal L}_S(K,\emptyset)$. We assert that ${\cal L}(K,U)$ is an upper subfamily of ${\cal L}_S$. For, if $A\in {\cal L}(K,U)$, $B\in{\cal L}_S$, and $A\subset B$,
then $B$ like $A$ must contain all members of $K$, so $B$ must belong to ${\cal L}(K,U)$.

In the other direction, suppose $\cal L$ is a learning space, and ${\cal L}^+$ is an upper subfamily of $\cal L$. Then, ${\cal L}^+$ can be represented as a fiber of a learning space, as follows. Form ${\cal L}'$ by adding to ${\cal L}$ the sets of the form $S\cup\{x\}$ where $S\in{\cal L}^+$ and $x\notin\cup\cal L$.
Project ${\cal L}'$ onto $\{x\}$; then ${\cal L}^+$ is the inverse image of $\{x\}$ under this projection.
\end{proof}

\begin{figure}[t]
\centering\includegraphics[width=3in]{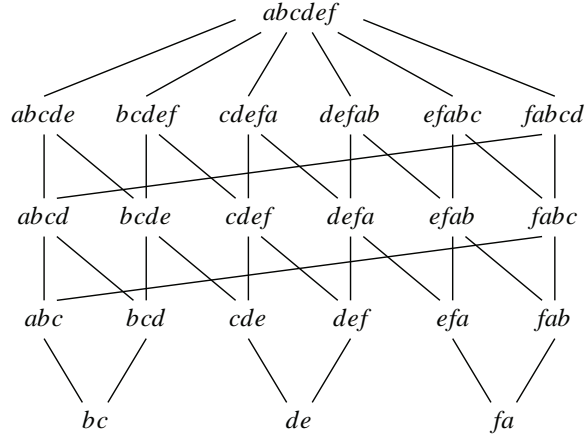}
\caption{A well-graded and union-closed set family that cannot be the fiber of a projected learning space.}
\label{fig:unfiber}
\end{figure}

Figure~\ref{fig:unfiber} shows a well-graded union-closed set family $\cal F$ that cannot be an upper subfamily ${\cal L}^+$ of a learning space ${\cal L}$. For $\cal L$ would have to contain a singleton as one of its states; by symmetry we can assume without loss of generality that this singleton is $\{a\}$. Then $\cal L$ would contain $\{a\}\cup\{d,e\}=\{a,d,e\}$. But $\{a,d,e\}$ is not in $\cal F$ while its subset $\{d,e\}$ is. Therefore, the family of fibers of projections of learning spaces is a proper subclass of the family of closed media.
\index{medium!closed}

\begin{theorem}
Let $\cal B$ be a given family of sets, the closure under union of which generates a family $\cal F$.
Then in time polynomial in the total size of the sets in $\cal B$ we can determine whether $\cal F$ is an upper subfamily ${\cal L}^+$ of a learning space ${\cal L}$, and if so find a collection of learning sequences defining $\cal L$.
\end{theorem}
\index{upper subfamily}
\index{worst case time complexity}
\index{learning sequence}

\begin{proof}
First, observe that we may test membership of a set $S$ in $\cal F$ in polynomial time, as $S\in\cal F$ if and only if $S=\cup\{T\in{\cal B}\mid T\subset S\}$. We define a set $S$ to be \emph{safe}
if, for any $T\in\cal F$, the union $S\cup T$ also belongs to $\cal F$. Equivalently, $S$ is safe if and only if, for any $T\in\cal B\cup\{\emptyset\}$, the union $S\cup T$ belongs to $\cal F$, so we may test safety in polynomial time.
If two sets $S$ and $T$ are safe, so is their union. If ${\cal F}\subset{\cal G}$, then ${\cal F}$ is an upper subfamily of $\cal G$ if and only if all sets in ${\cal G}\setminus{\cal F}$ are safe.
 
We say that a set $S$ is \emph{reachable} if we can find a sequence $\sigma_S$ of all the elements of $\cal S$ such that all prefixes of this sequence are safe. If $S$ and $T$ are both reachable, with $S\subset T$, then we may assume $\sigma_S$ forms an initial subsequence of $\sigma_T$, for if not the concatenation of $\sigma_S$ with the members of $\sigma_T$ not belonging to $\sigma_S$ forms another sequence on the elements of $T$ with the same property that all prefixes are reachable, by the union-closure of safe sets. Therefore, we may test reachability of $S$ in polynomial time, and find a sequence $\sigma_S$ for any reachable $S$, by the following greedy algorithm:
\index{greedy algorithm}

\begin{enumerate}
\item Initialize $\sigma$ to empty.
\item While $\sigma$ does not contain all elements of $S$:
\begin{enumerate}
\item If there exists $x$ in $S$ that can be added to $\sigma$ to form a longer safe prefix, do so.
\item Otherwise, terminate the algorithm and report that $S$ is not reachable.
\end{enumerate}
\item Return $\sigma_S=\sigma$.
\end{enumerate}

If $\cal F$ is an upper subfamily of a learning space $\cal L$, and $S\in\cal B$, then all sets in $\cal L$ must be safe, so $S$ is reachable via a learning sequence of $\cal L$ for which $S$ is a prefix. Therefore, if any set in $\cal B$ is not reachable, then $\cal F$ is not an upper subfamily of a learning space. On the other hand, if every set $S$ in $\cal B$ is reachable via a sequence $\sigma_S$,
then the unions of prefixes of these sequences form a learning space $\cal L$ containing $\cal B$ in which every set is safe; therefore $\cal B$ is an upper subfamily of this learning space $\cal L$.
The base of this learning space consists of certain prefixes of the sequences $\sigma_S$ for $S\in\cal B$, and we may use our concise representation algorithm to convert this base into a collection of learning sequences representing $\cal L$.
\end{proof}
\index{base}
\index{learning sequence}

This result gives us hope that there exists a simple combinatorial description of the set families that form upper subfamilies of learning spaces.
However, finding the \emph{smallest} learning space for which $\cal F$ is an upper subfamily is considerably more difficult.

\begin{theorem}
It is NP-complete to determine, for a given family of sets $\cal B$,  the closure under union of which generates a family $\cal F$, and for a given integer $K$, whether $\cal B$ is the upper subfamily of a learning space $\cal L$ with $|{\cal L}\setminus{\cal F}|\le K$.
\end{theorem}
\index{NP-completeness}

\begin{proof}
Note that, by our previous construction, $K$ need only be polynomial in the total size of $\cal B$.
Therefore, if there exists a suitably small $\cal L$, we may exhibit it by listing the additional sets
added to $\cal F$ to form $\cal L$, and test whether it correctly solves the problem by testing the safety of the additional sets, the union-closure of the added sets, and the reachability of each set in $\cal B$ by a sequence each prefix of which is a union of the added sets and of other sets already in $\cal B$.
Therefore, the problem is in NP.

To show NP-hardness, we reduce from the Vertex Cover problem, in which we are given a connected undirected graph $G$ and an integer $K$, and must find a set of $K$ or fewer vertices containing at least one endpoint of each edge of the graph. From an instance of Vertex Cover, we define $\cal B$ as containing a set of the two endpoints of each edge; the family $\cal F$ generated from $\cal B$ consists of all sets of two or more vertices from connected subgraphs of $G$.
Any learning space containing $\cal F$ must contain a subfamily of singleton sets for the vertices in a vertex cover of $G$, or some set in $\cal B$ would not be accessible, and conversely if $C$ is any vertex cover of $G$ we may form a learning space $\cal L$ with $|{\cal L}\setminus{\cal F}|=|C|$ by including a singleton set for each member of $C$.
\end{proof}
\index{vertex cover}

\subsection{Decomposing a Learning Space}

If ${\cal L}_1$ and ${\cal L}_2$ are two learning spaces on the same set of concepts, we may define a \emph{join} operation
combining the two into a single larger learning space:
$${\cal L}_1\sqcup {\cal L}_2 = \{ S_1 \cup S_2 \mid S_i\in{\cal L}_i \}.$$
This operation is commutative, associative, and \emph{idempotent} (that is, ${\cal L}\sqcup{\cal L}={\cal L}$ for any learning space $\cal L$); therefore it forms a \emph{semilattice} though not, in general, a medium.
\index{join!of two learning spaces}
\index{semilattice!of learning spaces}

The problem of computing the convex dimension of a learning space ${\cal L}$, and of representing ${\cal L}$ via a minimum number of learning sequences, can be viewed in this way as expressing $\cal L$ as the join of a small number of simpler learning spaces: if we define a \emph{totally ordered learning space} to be a learning space formed from a single learning sequence, then ${\cal L}$ has convex dimension at most $K$, and can be represented via $K$ learning sequences, if and only if it is the join of at most $K$ totally ordered learning spaces. The totally ordered learning spaces form the smallest subclass of learning spaces such that all learning spaces can be represented as joins of learning spaces in the subclass, for a learning space is irreducible for the join operation if and only if it is totally ordered.
\index{totally ordered learning space}
\index{learning sequence}

The algorithmic success of our learning sequence representation motivates us to seek more powerful forms of decompositions of learning spaces as joins of broader classes of learning spaces.
For, if $\cal L$ can be represented as a join of a small number of learning spaces ${\cal L}_i$, each of which in turn has an efficient algorithmic representation, then it seems likely that we may also perform
various algorithms efficiently for $\cal L$ itself, with an efficiency depending in part on the number $K$ of learning spaces used to represent $\cal L$. For instance, with such a representation, a base for $\cal L$ may be found as a subfamily of the union of bases of ${\cal L}_i$. But to use representions based on joins of special classes of learning spaces, we need to be able to find such representations efficiently. We have already shown that, if $\cal F$ is the family of totally ordered learning spaces, then a representation of any learning space as the join of a minimum number of members of $\cal F$ can be found efficiently. For which other families $\cal F$ is such efficient decomposition possible?
\index{join!of two learning spaces}

\begin{theorem}
It is NP-complete, given a base $\cal B$ of a learning space $\cal L$, and given an integer $K$, to determine whether $\cal L$ is the join of $K$ or fewer quasi-ordinal spaces.
\end{theorem}
\index{NP-completeness}
\index{quasi-ordinal space}

\begin{proof}
If $\cal L$ is the join of $K$ or fewer quasi-ordinal spaces, we may exhibit a partial order for each such space, test whether each base set of each such space belongs to $\cal L$, and test whether each base set of $\cal L$ belongs to at least one of these spaces; therefore, testing whether  $\cal L$ is the join of $K$ or fewer quasi-ordinal spaces can be done in NP.

To show NP-hardness, we reduce from the known NP-complete problem of graph edge coloring.
In this problem we are given an undirected graph $G$ and an integer $K$ and must assign $K$ distinct colors to the edges of $G$ in such a way that no two edges that share an endpoint have the same color.
From $G$ we form a learning space $\cal L$ with one concept per vertex $v$, and two concepts $\ell_e$ and $r_e$ per edge $e$. We include as base sets of $\cal L$ the singleton sets the singleton sets $\{\ell_e\}$ for each edge, the sets $\{ell_e,v\}$ for each edge $e$ having an endpoint $v$, and the sets $\{\ell_e,v,w,r_e\}$ for each edge $e$ having two endpoints $v$ and $w$.
\index{edge coloring}
\index{chromatic index}

Then, if ${\cal L}'\subset{\cal L}$ is a quasi-ordinal learning space containing a state $\{\ell_e,v,w,r_e\}$, the partial order defining ${\cal L'}$ must have $\ell_e$ (and no other concept) preceding $v$ and $w$.
Therefore, it can contain no other state $\{\ell_f,u,v,r_f\}$ for an edge $f$ sharing an endpoint with $e$. So, if $\cal L$ is the join of $K$ quasi-ordinal spaces, we can find a coloring of the edges of $G$ with $K$ colors, by coloring each edge according to which member of the join contains state  $\{\ell_e,v,w,r_e\}$.  Conversely, for any independent set of edges, we can find a quasi-ordinal subspace of $\cal L$ containing the states  $\{\ell_e,v,w,r_e\}$ for each edge in the independent set, so for each $K$-edge-coloring of $G$ we can find a representation of $\cal L$ as the join of $K$ quasi-ordinal spaces.
\index{join!of two learning spaces}

Since decomposition into $K$ quasi-ordinal spaces is in NP and a known NP-hard problem can be reduced to it, it is NP-complete.
\end{proof}

The same proof shows that the problem remains hard when $K$ is fixed to any constant greater than two. A similar reduction (omitting the concepts $\ell_e$) shows that it is NP-complete to represent a learning space as a join of three or more \emph{hierarchies}, where a hierarchy is defined as a quasi-ordinal learning space for an order the Hasse diagram of which has at most one outgoing edge per concept. A hierarchy can be characterized as a learning space each two base sets of which are either disjoint or nested, so one can test whether a learning space $\cal L$ is a join of two hierarchies by forming a graph in which we connect two base sets of $\cal L$ by an edge whenever they do not satisfy this condition and testing whether this graph is bipartite; we omit the details. We have not yet determined whether it is possible to test efficiently whether a given learning space is the join of two quasi-ordinal spaces.
\index{hierarchy}
\index{Hasse diagram}

\subsection{Antimatroids as Algebras}

We have described learning spaces (or antimatroids) here as union-closed accessible families of sets, and they are also commonly formalized as prefix-closed formal languages satisfying a certain exchange axiom \citep{KorLovSch-91}. Media theory \citep{falma02,eppst07b} leads to alternative formalizations involving geometric graphs and finite automata, and previous work on drawing antimatroids \citep{eppst06} leads to a characterization in terms of arrangements of translated orthants. But it turns out that antimatroids also have a fairly natural algebraic characterization. That is, we view antimatroids in terms of the algebraic properties of their union operation rather than as systems of explicit sets and elements. Perhaps this is not surprising, as the first work on antimatroids \citep{dilworth40} characterized them algebraically in terms of lattices, but we use semilattices instead.
\index{formal language}
\index{prefix}
\index{exchange axiom}
\index{geometric graph}
\index{finite automaton}
\index{graph drawing}
\index{orthant}
\index{lattice}
\index{semilattice}
\index{Korte, B.}
\index{Lov{\'a}sz, L.}
\index{Schrader, R.}
\index{Falmagne, J.-Cl.}
\index{Ovchinnikov, S.}
\index{Eppstein, D.}
\index{Dilworth, R.P.}

The model for this is an algebraic description of semilattices and quasi-ordinal spaces (equivalently, finite distributive lattices or finite lattices of sets). A \emph{monoid} consists of an associative binary operation (such as multiplication or string concatenation or function composition) on a set of objects, together with an identity object for the operation (such as, respectively, 1 or the empty string or the identity function); here we write the monoid operation as multiplication. A \emph{semilattice} is a monoid that is commutative ($xy=yx$ for all $x,y$) and idempotent ($x^2=x$ for all $x$). The \emph{divisibility} relation on a semilattice ($x|y$ if, for some $z$, $xz=y$) forms a partial order, and the semilattice operation can be interpreted as a join or least upper bound operation for that partial order. By analogy with ring theory we can define an \emph{irreducible} object in a semilattice to be an $x$ such that if $x=yz$ then $x=y$ or $x=z$, and we can define a \emph{prime} object in a semilattice to be an $x$ such that if $x|yz$ then $x|y$ or $x|z$. Lattices are normally defined in terms of two compatible semilattices, one for joins and one for meets, but as we see below the concepts of irreducibility and primality allow us to define finite lattices of sets in terms of the join operation only.
\index{quasi-ordinal space}
\index{lattice!of sets}
\index{lattice!distributive}
\index{monoid}
\index{semilattice}
\index{divisibility!in a semilattice}
\index{irreducibility!in a semilattice}
\index{primality!in a semilattice}

\begin{lemma}
In any semilattice, if $x$ is prime then $x$ is irreducible.
\end{lemma}

\begin{proof}
Let $x$ be prime, and let $x=yz$. Since $x|x=yz$, by primality $x|y$ or $x|z$, and without loss of generality $x|y$, but $y|x$ so $y=x$.
\end{proof}

\begin{lemma}
\label{lem:prodirrediv}
In a finite semilattice every object is the product of its irreducible divisors.
\end{lemma}

\begin{proof}
The result is trivial if an object $x$ is irreducible, and otherwise if $x=yz$ the result follows by induction on the number of irreducible factors of $y$ and $z$.
\end{proof}

\begin{theorem}
A finite semilattice can be represented as the union operation of a quasi-ordinal space if and only if every irreducible is prime.
\end{theorem}
\index{quasi-ordinal space}

\begin{proof}
In a quasi-ordinal space, or more generally an antimatroid, the irreducibles are the sets of a base of the space, and as we have seen for a quasi-ordinal space these sets are of the form $B_a=\{b\mid b\le a\}$ for some element $a$ of the partial order defining the space. In the semilattice for the union operation of the space, divisibility is the subset relation, $B_x\subset S$ if and only if $a\in S$, and if $a\in S$ and $S=T\cup U$ then $a\in T$ or $a\in U$, so the definition of primality follows.

In the other direction, suppose in a semilattice every irreducible is prime, and represent every semilattice object $x$ by the set $P_x$ of primes dividing $x$. Then by Lemma~\ref{lem:prodirrediv}, $x$ is the product of $P_x$, so each $P_x$ corresponds to a single object $x$.  If $x=yz$, $P_x=P_y\cup P_z$ by primality, so the semilattice operation is represented by the union operation on the sets $P_x$. Every set $P_x$ is a lower set on the restriction of the divisibility order of the semilattice to the primes, and if $P$ is a any lower set in this order having product $x$ then $P=P_x$, so the original semilattice is isomorphic to the quasi-ordinal space generated by the divisibility order on the primes.
\end{proof}

Antimatroids lie between semilattices and quasi-ordinal spaces in generality, so it seems natural to axiomatize them similarly. However, it doesn't appear to work to use primes and irreducibles as with lattices: that axiomatization corresponds set elements with the minimum lattice object containing that element, but in antimatroids there may be no unique minimum object. Instead it works better to turn things upside down, and correspond elements with the maximum object not containing the element.

\begin{lemma}
In a finite semilattice, $p$ is irreducible if and only if it has a single \emph{predecessor} $q$, such that $q|p$ and, for any $x$, if $x|p$ then $x=p$ or $x|q$.
\end{lemma}
\index{predecessor}

\begin{proof}
If $q_0$ and $x_0$ violate this definition, as do $q_i=q_{i-1}x_{i-1}$ and some $x_i$ for each $i>0$, the values of $q_i$ would form an infinite ascending chain of proper divisors of $p$ violating the assumption of finiteness. Conversely, if $p$ has a single predecessor $q$, it must be irreducible, as any product of its proper divisors would also divide $q$ and therefore be unequal to $p$.
\end{proof}

By analogy, in any semilattice, we define $s$ to be \emph{singular} iff it has a single \emph{successor} $t$, such that $s|t$ and, for any $x$, if $s|x$ then $s=x$ or $t|x$. We identify any object $x$ in the semilattice with the set $N(x)$ of singular objects that $x$ does not divide.

\begin{lemma}
For any $x$ and $y$, $N(xy)=N(x)\cup N(y)$.
\end{lemma}

\begin{proof}
If $s\in N(xy)$, then $xy\notdiv s$. But in any semilattice, for any $s$, if $x|s$ and $y|s$ then $xy|s$, so we can conclude that $x\notdiv s$ or $y\notdiv s$, and therefore $s\in N(x)$ or $s\in N(y)$.
Conversely if $s\in N(x)$, so $x\notdiv s$, then $xy\notdiv s$ and $s\in N(xy)$.
\end{proof}

\begin{theorem}
\label{thm:Nsemilat}
Any finite semilattice is isomorphic to the semilattice formed by the union operation on the sets $N(x)$.
\end{theorem}

\begin{proof}
From the previous lemma, it remains only to show that for any $x\ne y$, $N(x)\ne N(y)$. Suppose to the contrary that $S=\{x\mid N(x)=T\}$ has $|S|>1$ for some set $T$ of singular objects. By the previous lemma, $S$ is closed under unions, so it contains a unique maximal object, and we may choose from $S$ some two objects $x\ne y$ with $x|y$. Let $U=\{z\mid x|z\mbox{ and }y\notdiv z\}$,
and let $u$ be a maximal object in $U$; $U$ is nonempty as it contains $x$ and a maximal $u$ exists by finiteness. Then $uy$ must be the unique successor to $u$, for if $u$ had a proper multiple $w$ not divisible by $uy$, then $w$ would be a multiple of $x$ and nonmultiple of $y$ that is larger than $u$ contradicting the maximality of $u$. Therefore, $u$ is singular, and $N(y)$ contains $u$ but $N(x)$ doesn't, contradicting the assumption that $N(x)=N(y)$.
\end{proof}

\begin{figure}[t]
\centering\includegraphics[scale=0.45]{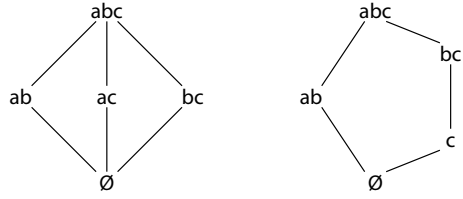}
\caption{Two semilattices that are not antimatroids.}
\label{fig:slnam}
\end{figure}

In some sense the sets $N(x)$ form a minimal representation by sets for any finite semilattice. It remains to find conditions on the semilattice forcing these sets to be accessible. There are many possible conditions that do so; the simplest we have found is the following. We define an \emph{equalizing pair} $x,y$ to be a pair of objects $x|y$ such that there exist $a$ and $b$ with $xa\ne xb$ and $xa\ne ya=yb\ne xb$. We say that a finite semilattice has \emph{separated equalizers} if, for each equalizing pair $x,y$, there exists $z$ with $x|z|y$ and $x\ne z\ne y$. This definition is motivated by the two semilattices shown in Figure~\ref{fig:slnam}. In the left semilattice, the pair $\emptyset,ab$ is equalizing for $ac$ and $bc$, while in the right semilattice, the pair $\emptyset,ab$ is equalizing for $c$ and $bc$; however, in each case there is no $z$ between $\emptyset$ and $ab$. Thus, from the lemma below, the property of having separated equalizers distinguishes these two semilattices from antimatroids.
\index{antimatroid}

\begin{lemma}
Every antimatroid has separated equalizers.
\end{lemma}

\begin{proof}
Let $x,y$ be an equalizing pair for $a,b$ in an antimatroid. We can assume without loss of generality that $xa\notdiv xb$. Then, in order for $xa\ne xb$ but $ya=yb$, $y\setminus x$ must be a superset of $xa\setminus xb$. It must be a proper superset of this set, else we would have $ya=xa$. Therefore, $|y\setminus x|>1$ and we can find $z$ between $x$ and $y$ using the chain property for antimatroids.
\end{proof}

\begin{lemma}
\label{lem:depent}
In a semilattice with separated equalizers, suppose that there exist objects $a$, $b$, $c$, and $d$, with $a|b$, $a|c|d$, and $bc=bd$. Then there exists another object $x$, with $a|x|b$, such that either $xc=d$, or  $xc$ and $d$ are incomparable.
\end{lemma}

\begin{proof}
An equivalent description of the premise of the lemma is that $a,b$ form an equalizing pair for some $c$ and $d$ with $c|d$. We use induction on the length of the longest chain of objects between $a$ and $b$. Because $a,b$ form an equalizing pair for $c,d$, there must exist $x$ between $a$ and $b$.
If $xc$ is a proper divisor of $d$, then $x,b$ form an equalizing pair for $xc$ and $d$ and the result follows by induction.  If $xc=d$, we have proven the lemma. If $d|xc$, then $a,x$ form an equalizing pair for $c$ and $d$ and the result follows by induction. In the remaining case, $xc$ and $d$ are incomparable.
\end{proof}

\begin{figure}[t]
\centering\includegraphics[scale=0.45]{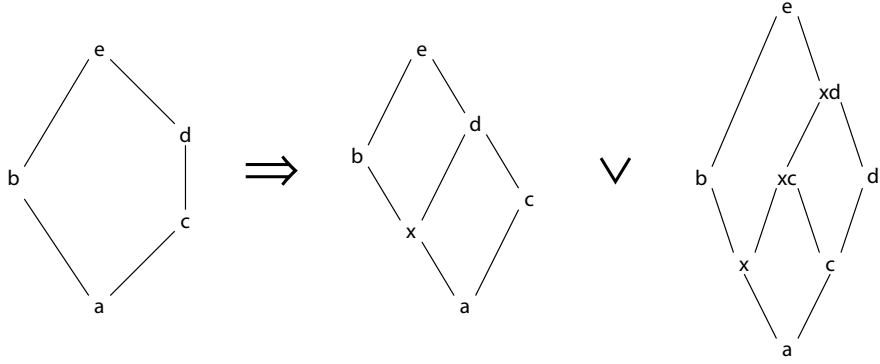}
\caption{Graphical explanation of Lemma~\ref{lem:depent}.}
\label{fig:depent}
\end{figure}

A graphical explanation of Lemma~\ref{lem:depent} is shown in Figure~\ref{fig:depent}.

\begin{lemma}
\label{lem:sing-is-max}
In a semilattice with separated equalizers, if $p$ is irreducible with predecessor $q$, $s$ is singular with successor $t$, $q|s$, and $p\notdiv s$, then $t=ps$.
\end{lemma}

\begin{proof}
By singularity of $s$, $ps=pt$, so if $t\ne ps$ then $q,p$ form an equalizing pair for $s,t$. But if $p$ and $q$ are separated by $z$, $q|z|p$, then $q$ could not be the predecessor of $p$.
\end{proof}

\begin{lemma}
\label{lem:irredsingeq}
In a semilattice with separated equalizers, let $p$ be irreducible with predecessor $q$,
$s$ be singular with successor $t$, $q|s$, and $p\notdiv s$. Then there can be no $x$ with $q|x$, $p\notdiv x$, and $s$ incomparable to $x$.
\end{lemma}

\begin{proof}
Suppose to the contrary that $x$ exists.
Then by singularity of $s$, $t|xs$. If $t=xs$, then $q,p$ would be an equalizing pair for $s$ and $x$, and the existence of $q|x|p$ would violate the assumption that $q$ is $p$'s predecessor.
Otherwise, $t$ is a proper divisor of $xs$, so $q,x$ is an equalizing pair for $s$ and $t$.
By Lemma~\ref{lem:depent}, there exists $x'$ with $q|x'|x$ where either $sx'=t$ or $x's$  and $t$ are incomparable;  $x'$ must be incomparable to $s$.
By repeating this process we can find an infinite descending sequence of $x$, $x'$, $x''$, etc., violating the assumption of finiteness of the semilattice, so $x$ cannot exist.
\end{proof}

\begin{lemma}
\label{lem:irred-single}
In a semilattice with separated equalizers, let $p$ be irreducible with predecessor $q$.
Then $|N(p)\setminus N(q)|=1$.
\end{lemma}

\begin{proof}
$|N(p)\setminus N(q)|\ge 1$ by Theorem~\ref{thm:Nsemilat}.
By Lemma~\ref{lem:irredsingeq}, any two members of $N(p)\setminus N(q)$ must be comparable, but by Lemma~\ref{lem:sing-is-max}, any two members must be incomparable.
Therefore $|N(p)\setminus N(q)|\le 1$.
\end{proof}
 
\begin{theorem}
A finite semilattice can be represented as an antimatroid if and only if it has separated equalizers.
\end{theorem}

\begin{proof}
For any x in such a semilattice, there is a $y|x$ with $|N(x)\setminus N(y)|=1$: represent $x$ as a minimal product of irreducibles, and form $y$ by replacing one of the irreducibles in this product by its predecessor. Therefore the sets $N(x)$ are accessible as well as union-closed, and this set family forms an antimatroid. Conversely we have seen that any antimatroid forms a semilattice that has separated equalizers.
\end{proof}

\subsection{Different Definitions of Dimension}

Our algorithms for learning spaces have been centered around the concept of \emph{convex dimension}, the minimum number ${\rm dim}_C({\cal L})$ of learning sequences needed to define the learning space $\cal L$.
However, there are several other natural concepts of dimension for learning spaces.
We may define the \emph{base dimension} ${\rm dim}_B({\cal L})$of a learning space $\cal L$ to be the cardinality of its base.
The \emph{lattice dimension} ${\rm dim}_\Zee({\cal L})$ \citep{eppst05} is the minimum dimension $d$ of an integer lattice $\Zee^d$ into which the states of $\cal L$ may be embedded, in such a way that the $L_1$ distance between the embeddings of two states equals the cardinality of their symmetric difference; like the convex dimension it can be calculated efficiently by an algorithm based on maximum matching in an associated bipartite graph. And, the \emph{order dimension} ${\rm dim}_\le({\cal L})$
is the minimum dimension $d$ of a Euclidean space $\Ree^d$ into which the states of $\cal L$ may be embedded, in such a way that, for two states $S$ and $T$, $S\subseteq T$ if and only if the coordinates of $S$ are all less than or equal to the corresponding coordinates of $T$. In some sense the order dimension is very closely related to the convex dimension, as both are defined as the minimum number of sequences of items needed to define the learning space, but in the case of convex dimension the sequences are of elements of $\cup L$ while in the case of order dimension the sequences (formed by each coordinate of the embedding) are of states of $\cal L$. We may also view the cardinality $n=|{\cup\cal L}|$ as a dimension: it is the \emph{isometric dimension} of $\cal L$, that is, the least dimension of a hypercube into which $\cal L$ can be isometrically embedded.
\index{dimension}
\index{convex dimension}
\index{base dimension}
\index{lattice dimension}
\index{order dimension}
\index{isometric dimension}
\index{Euclidean space}
\index{Eppstein, D.}
\index{base}
\index{learning sequence}

For any learning space $\cal L$, these different quantities satisfy the following inequalities and relations:

\begin{itemize}
\item ${\rm dim}_C({\cal L})\le{\rm dim}_B({\cal L})$. This follows as we may represent $\cal L$ using a separate learning sequence for each base set.

\item $n\le{\rm dim}_B({\cal L})$. Each element of $\cup\cal L$ must be the removable element of at least one base set.

\item ${\rm dim}_B({\cal L})\le{\rm dim}_C({\cal L})\cdot n$, as each learning sequence in a representation of $\cal L$ by learning sequences can contribute at most $n$ base sets.

\item ${\rm dim}_C({\cal L})\le{n\choose\lfloor n/2\rfloor}=O(2^n/\sqrt n)$, by our characterization of ${\rm dim}_C({\cal L})$ as the size of the largest antichain in the base, and Sperner's Theorem bounding the size of an antichain in any family of sets.

\item ${\rm dim}_\le({\cal L})\le{\rm dim}_C({\cal L})$ \citep{KorLovSch-91}. This can be seen via the embedding into $\Ree^{{\rm dim}_C({\cal L})}$ in which we map $S$ to ${\rm mex}(S)$, for this embedding satisfies the requirements of the definition of order dimension.
\index{Korte, B.}
\index{Lov{\'a}sz, L.}
\index{Schrader, R.}

\item ${\rm dim}_\le({\cal L})=2$ if and only if ${\rm dim}_C({\cal L})=2$, from our work on drawing learning spaces \citep{eppst06}.
\index{Eppstein, D.}
\index{graph drawing}

\item ${\rm dim}_\le({\cal L})\le{\rm dim}_\Zee({\cal L})$: in any lattice embedding of $\cal L$, all states must be mapped to a single orthant of the lattice in order to satisfy the union closure property of $\cal L$, so the lattice embedding must again satisfy the requirements of the definition of order dimension.
\index{orthant}

\item ${\rm dim}_\Zee({\cal L})\le n$, as the characteristic function embeds $\cal L$, or more generally any family of sets on the elements of $\cal L$, into the subset $\{0,1\}^{n}$ in such a way that $L_1$ distance equals symmetric difference cardinality.
\index{characteristic function}

\begin{figure}[t]
\centering\includegraphics[width=2.5in]{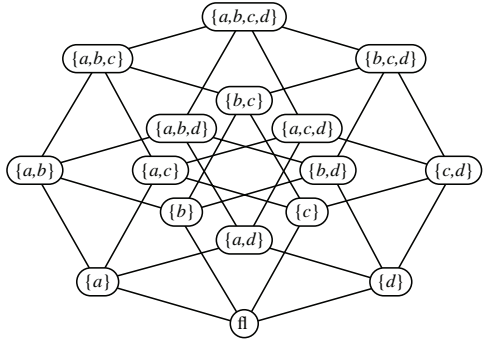}
\caption{A powerset on four elements.}
\label{fig:4cube}
\end{figure}

\item ${\rm dim}_C({\cal L})=O(n^{{\rm dim}_\Zee({\cal L})-1})$.
This follows from the fact that a lattice embedding of $\cal L$ must lie within a product of intervals $[0,n-1]$ and the fact that no two members of an antichain within this product can share all but one coordinate.
\end{itemize}

\noindent
We now describe examples of learning spaces that are extremal for some of these inequalities.

\begin{description}
\item[A chain.] \hfill\\
A learning space defined from a single learning sequence has
${\rm dim}_C({\cal L})={\rm dim}_\le({\cal L})={\rm dim}_\Zee({\cal L})=1$ but
${\rm dim}_B({\cal L})={\rm dim}_C({\cal L})\cdot n=n$.
\index{learning sequence}

\item[A powerset.] \hfill\\
The family of all subsets of an $n$-element set (Figure~\ref{fig:4cube}) has ${\rm dim}_C({\cal L})={\rm dim}_\le({\cal L})={\rm dim}_\Zee({\cal L})={\rm dim}_B({\cal L})=n$.
\index{power set}

\begin{figure}[t]
\centering\includegraphics[width=4in]{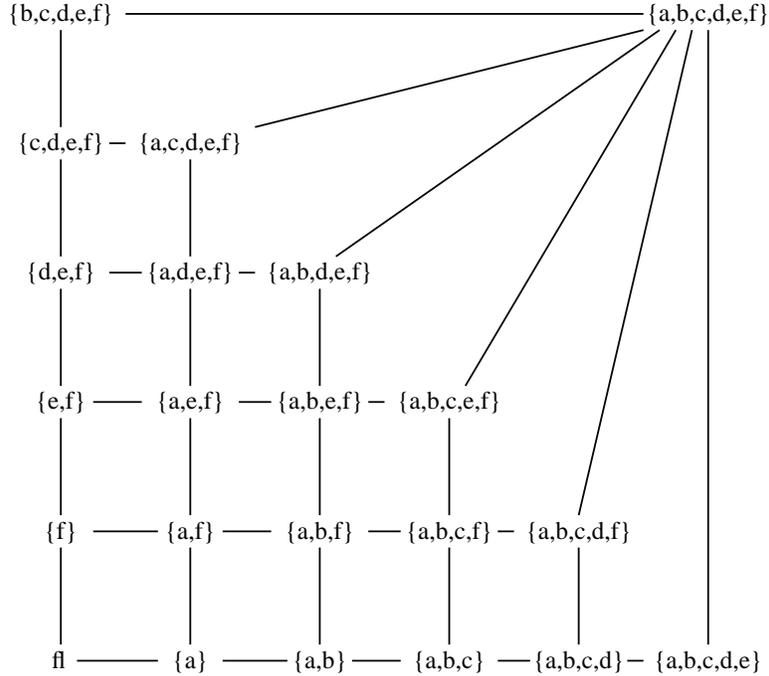}
\caption{The learning space defined by a learning sequence and its reverse. Figure from \citet{eppst06}.}
\label{fig:prefixsuffix}
\end{figure}

\item[A learning sequence and its reverse.] \hfill\\
The learning space defined by two learning sequences, one the reverse of the other
(Figure~\ref{fig:prefixsuffix}) has
${\rm dim}_C({\cal L})={\rm dim}_\le({\cal L})=2$ but
${\rm dim}_\Zee({\cal L})=n$ and ${\rm dim}_B({\cal L})=2n-2$.
\index{Eppstein, D.}

\item[A learning space with a large base.] \hfill\\
Let $D$ be a set of $n$ elements, and $x$ be a designated element from $D$.
Define $\cal L$ to consist of the sets that either do not contain $x$, or
contain at least $\lfloor n-1)/2\rfloor$ elements.
Then $\cal L$ is accessible, as for any set $S\in L$, if $x\in S$ then $S\setminus\{x\}\in\cal L$
while if $x\notin S$ then all subsets of $S$ are in $\cal L$.
$\cal L$ is also closed under unions, so it forms  a learning space.
The base of $\cal L$ consists of $\{x\}$ together with all subsets of exactly
$\lfloor (|D|-1)/2\rfloor$ elements of $D\setminus\{x\}$; the subset of the base formed by omitting $\{x\}$ is an antichain.
Therefore ${\rm dim}_\le({\cal L})\le{\rm dim}_\Zee({\cal L})=n$,
but ${\rm dim}_B({\cal L})=1+{n-1 \choose \lfloor (n-1)/2 \rfloor}$ and
${\rm dim}_C({\cal L})={n-1 \choose \lfloor (n-1)/2 \rfloor}$, matching to within a constant factor the $O(2^n/\sqrt n)$ upper bound on ${\rm dim}_C({\cal L})$.

\begin{figure}[t]
\centering\includegraphics[width=4in]{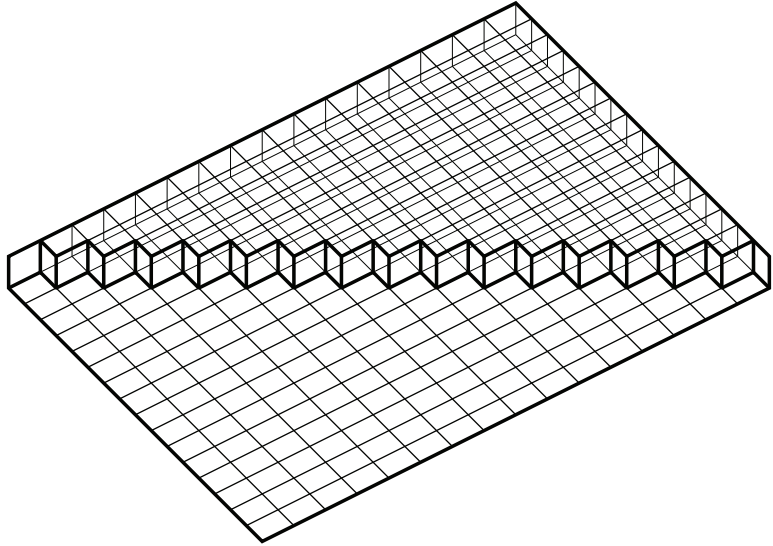}
\caption{A learning space with large convex dimension but low lattice dimension.}
\label{fig:zigzag}
\end{figure}

\item[A three-dimensional zigzag.] \hfill\\
Let $N$ be a given number, and let $Z$ consist of the three-dimensional integer lattice points with coordinates $0\le x,y<N$, $0\le z\le 1$, and such that, if $z=1$, then $x+y+z\ge N$ (Figure~\ref{fig:zigzag}).
The semilattice of coordinatewise maximization on $Z$ can be represented as a learning space $\cal L$ with $n=|{\cup\cal L}|=2N-1$. For this space,
${\rm dim}_\le({\cal L})={\rm dim}_\Zee({\cal L})=3$
but ${\rm dim}_C({\cal L})=N$, as there is an antichain in the base consisting of all minimal points with $z=1$. Similar examples in higher dimensions show that, in general, ${\rm dim}_C({\cal L})$ can be lower bounded by $\Omega(n^{{\rm dim}_\Zee({\cal L})-2})$,
nearly matching our $O(n^{{\rm dim}_\Zee({\cal L})-1})$ upper bound.
\end{description}

It would be of interest to determine the algorithmic complexity of calculating the order dimension of a learning space. For arbitrary partial orders, calculating the order dimension is NP-complete \citep{Yan-SIADM-82} but it is unclear whether the reduction proving this can be made to apply to learning spaces.
\index{order dimension}
\index{worst case time complexity}
\index{Yannakakis, M.}

\section{Future Work}

Although we have made significant progress in learning space implementation, we believe there is plenty of scope for additional investigation, particularly on the following topics:

\begin{description}
\item[Visualization of learning spaces.] \hfill\\
Quasi-ordinal spaces may be visualized by drawing their Hasse diagrams as graphs, but this technique does not work so well for more general learning spaces.
In earlier work \citep{eppst04,eppst06} we found algorithms for drawing the state-transition diagrams of learning spaces and more general media, however these work best when the space being drawn has a well-behaved embedding into a low-dimensional Euclidean space. Can we generalize these drawing approaches to learning spaces with higher convex dimension? Is there a possibility of a hybrid approach that  draws portions of a learning space as a Hasse diagram on concepts and resorts to the more complex state space only when necessary?
\index{graph drawing}
\index{state-transition diagram}
\index{Euclidean space}
\index{Hasse diagram}
\index{Eppstein, D.}

\item[Counting states.] \hfill\\
Is there an efficient method for counting or estimating the number of states in a learning space, without taking the time to generate all states? This would have implications for our ability to set sample sizes appropriately in ALEKS's knowledge assessment algorithms, as well as for calculating more accurate priori probability distributions on projected states when using projections of learning spaces to speed up the assessment algorithm. From past experience with similar combinatorial counting problems \citep[e.g.,][]{JerSinVig-STOC-01} we expect that the complexity of a randomized approximation scheme for the counting problem is likely equivalent to the complexity of sampling states from the learning space uniformly at random, which seems to be of interest independently.
\index{state generation}
\index{randomized algorithm}
\index{approximation algorithm}
\index{Jerrum, M.}
\index{Sinclair, A.}
\index{Vigoda, E.}

\item[Inference of error rates.] \hfill\\
ALEKS  currently assumes fixed rates of careless mistakes and lucky guesses, which it uses in its calculation of likelihoods that a student knows each concept in a learning space. But one could also envision a more sophisticated Bayesian assessment procedure that treats the chance of careless errors as a variable with an a priori probability distribution, and attempts to infer a maximum likelihood value of this variable based on the student's answers. Such a procedure would likely be based on an EM-algorithm approach in which one alternates applications of the likelihood calculation described here with an algorithm for estimating the careless error probability given the calculated likelihoods, and would allow the system to better fit its model to each student's performance. However the details of such an approach still need to be worked out.
\index{careless mistake}
\index{lucky guess}
\index{Bayesian statistics}
\index{EM algorithm}

\item[Reconciliation of expert opinions.] \hfill\\
Along with its applications to concise representation of media, the join may be useful for a problem arising when constructing learning spaces from the answers of experts \citep{dowling:93a}: two different experts may give quite different answers when asked what they believe about the prerequisite structure of a given set of concepts, leading to quite different learning spaces on those concepts. The construction procedure of \citet{dowling:93a} involves asking experts a series of questions about whether feasible knowledge states can exist with certain combinations of concepts, but the answers to these questions have only been found reliable when the combinations involve at most two concepts at a time; the learning spaces generated by limiting the questioning to such combinations are necessarily quasi-ordinal.  To reliably generate more complex learning spaces, it seems necessary to combine the results from questioning multiple experts.
The join provides a mathematical mechanism for reconciling those answers and finding a common learning space containing as states any set of concepts believed to form a state by any of the experts,
but the learning spaces constructed in this way are likely to be much larger than necessary.
More research is needed on methods for combining information from multiple experts to generate learning spaces of size comparable to the space that would be constructed by questioning a single expert, while simultaneously taking advantage of the multiplicity of experts to generate spaces that more accurately model the students' knowledge.
\index{join!of two learning spaces}
\index{semilattice!of learning spaces}
\index{expert knowledge}
\index{knowledge engineering}
\index{Dowling, C.E.}

\item[Even faster state space generation.] \hfill\\
Our algorithm for generating the states in an $n$-concept quasi-ordinal space takes time $O(n)$ per generated state, without assumption, and may often be faster, while a similarly fast time bound of $O(k)$ per generated state for learning spaces generated by $k$ learning sequences can only  be shown with an additional assumption of constant time bitvector operations for maintaining and updating the ${\rm mex}$ values of the generated states. Additionally, we have briefly described an algorithm for listing all states in the fiber ${\cal L}(K,U)$ of a projection of a learning space, given beliefs that a student knows the concepts in $K$ and does not know the concepts in $U$, that do not match these efficiencies. Can these state generation algorithms be improved to the point where more efficient worst case guarantees on their performance can be proven?
\index{state generation}
\index{minimum excluded element}
\index{worst case time complexity}

\item[Faster upper fringe construction.] \hfill\\
The algorithm we implemented for our construction of the family of sets that can be added to a learning space to form new larger learning spaces involves generating all states of the learning space. However, there may be many fewer fringe sets than there are states.  In some sense, the base of a learning space consists of the minimal sets in the space, while the outer fringe consists of the maximal sets not in the space.  Thus, it is plausible that one could adapt hypergraph transversal algorithms \citep{FreKha-Algs-96}, which can be used to convert minimal sets in a family to maximal sets not in a family for certain other types of set families, to the purpose of finding the outer fringe of a learning space in time pseudopolynomial in the number of sets in the outer fringe. Such a result would also have implications for the computational complexity of inferring a learning space from questions asked of an expert \citep{dowling:93a}. However, we have not worked out the details of such an efficient  algorithm for listing upper fringe states.
\index{fringe!of a learning space}
\index{Michael L. Fredman}
\index{Khachiyan, L.}
\index{Dowling, C.E.}

\item[Structure of the family of learning spaces.] \hfill\\
We know from the work of \citet{Thi-01} that, when one learning space forms a subfamily or superfamily of the other, we can find a shortest path from one to the other by adding and removing states, such that each set family in this shortest path is also a learning space. That is, the family of learning spaces has a chain property similar to that of individual learning spaces. This fact motivates the calculation of the fringes of a learning space, as the sets in
the fringe represent potential neighbors in such paths. We also know that the family of learning spaces on a given domain forms a semilattice under the join operation, which is however not the same as simple union of set families. And we know that the family of learning spaces is not in general well-graded, so it does not form a medium under operations that add and remove sets. What other structure does the family of learning spaces have, and how can that structure help us quickly adapt a learning space to changing information about the possible knowledge states of students?
\index{Thi{\'e}ry, N.}
\index{adaptation}

\item[Question selection strategy.] \hfill\\
We have only indirectly addressed the issue of which question to ask the student next, in the assessment procedure, after likelihoods of each concept have been calculated.
As currently implemented, ALEKS selects the question with likelihood closest to 50\% of being known, but that may not be the optimal selection strategy. It seems likely that a somewhat better strategy would be to select the question with likelihood closest to 50\% of being answered correctly; due to the different rates of careless errors and lucky guesses this strategy differs from the currently implemented one, but it also depends on having an accurate estimate of the student's careless error rate, which the current assessment procedure does not supply. Also, if there are multiple questions with similar likelihoods, it may be best not to choose the one with likelihood closest to 50\%, but instead to perform some lookahead in the sequence of questions, and ask a question such that whichever answer is given will again lead to a situation where some question has likelihood close to 50\%. The effect of an improved question selection strategy could be to reduce the number of questions needed to assess each student's knowledge, over and above the reduction afforded by more accurately defining the learning space on which the assessment is based. In addition, the outcome of a question may not actually be binary: ``don't know'' may be treated differently than an incorrect answer, and the nature of the errors in an incorrect answer may yield some insight about the student's knowledge. It seems likely that these brief initial observations could be significantly expanded with more thought.
\end{description}

\section{Conclusions}

We have shown that a computer representation of learning spaces by learning sequences can approach the efficiency of the existing quasi-ordinal space representation for ALEKS's knowledge assessment algorithms, while allowing a broader class of learning spaces that may more easily be adapted by adding and removing states. We believe that the algorithms described here are sufficiently detailed and efficient to be suitable for implementation within ALEKS.

We have also performed theoretical investigations concerning learning spaces. We have described how to recognize spaces formed from a learning space when we assume certain facts about the state of a student's knowledge, and we have investigated the algorithmic complexity of recognizing learning spaces that can be decomposed into joins of a small number of simpler spaces. We have investigated alternative mathematical representations of learning spaces, and we have compared the convex dimension fundamental to our computer representation to several other important numerical measures of a learning space's size.

Finally, we have identified multiple areas where more research may lead to additional practical algorithms or theoretical insights concerning learning spaces.

\ifstandalone

\raggedright
\bibliographystyle{plainnat}
\bibliography{grand}

\begin{thebibliography}{21}
\providecommand{\natexlab}[1]{#1}
\providecommand{\url}[1]{\texttt{#1}}
\expandafter\ifx\csname urlstyle\endcsname\relax
  \providecommand{\doi}[1]{doi: #1}\else
  \providecommand{\doi}{doi: \begingroup \urlstyle{rm}\Url}\fi

\bibitem[Avis and Fukuda(1996)]{avis96}
D.~Avis and K.~Fukuda.
\newblock {Reverse search for enumeration}.
\newblock \emph{Discrete Applied Mathematics}, 65:\penalty0 21--46, 1996.

\bibitem[Birkhoff(1937)]{birkh37}
G.~Birkhoff.
\newblock {Rings of sets}.
\newblock \emph{Duke Mathematical Journal}, 3:\penalty0 443--454, 1937.

\bibitem[Boyd and Faigle(1990)]{BoyFai-DAM-90}
E.A. Boyd and U.~Faigle.
\newblock {An algorithmic characterization of antimatroids}.
\newblock \emph{Discrete Applied Mathematics}, 28:\penalty0 197--205, 1990.

\bibitem[Dilworth(1940)]{dilworth40}
R.P. Dilworth.
\newblock {Lattices with unique irreducible decompositions}.
\newblock \emph{Annals of Mathematics}, 41:\penalty0 771--777, 1940.

\bibitem[Doble et~al.(2001)Doble, Doignon, Falmagne, and Fishburn]{doble01a}
C.W. Doble, J.-P. Doignon, J.-Cl. Falmagne, and P.C. Fishburn.
\newblock {Almost connected orders}.
\newblock \emph{Order}, 18\penalty0 (4):\penalty0 295--311, 2001.

\bibitem[Doignon and Falmagne(1999)]{doign99}
J.-P. Doignon and J.-Cl. Falmagne.
\newblock \emph{{Knowledge Spaces}}.
\newblock Springer-Verlag, Berlin, Heidelberg, and New York, 1999.

\bibitem[Dowling(1993)]{dowling:93a}
C.E. Dowling.
\newblock {Applying the basis of a knowledge space for controlling the
  questioning of an expert}.
\newblock \emph{Journal of Mathematical Psychology}, 37:\penalty0 21--48, 1993.

\bibitem[Eppstein(2005{\natexlab{a}})]{eppst04}
D.~Eppstein.
\newblock {Algorithms for drawing media}.
\newblock In \emph{Graph Drawing: 12th International Symposium, GD 2004, New
  York, NY, USA, September 29--October 2, 2004}, volume 3383 of \emph{Lecture
  Notes in Computer Science}, pages 173--183, Berlin, Heidelberg, and New York,
  2005{\natexlab{a}}. Springer-Verlag.

\bibitem[Eppstein(2005{\natexlab{b}})]{eppst05}
D.~Eppstein.
\newblock {The lattice dimension of a graph}.
\newblock \emph{European Journal of Combinatorics}, 26(6):\penalty0 585--592,
  2005{\natexlab{b}}.

\bibitem[Eppstein(2006)]{eppst06}
D.~Eppstein.
\newblock {Upright-quad drawing of $st$-planar learning spaces}.
\newblock In \emph{Graph Drawing: 14th International Symposium, GD 2006,
  Karlsruhe, Germany, September 18--20, 2006}, Lecture Notes in Computer
  Science, Berlin, Heidelberg, and New York, 2006. Springer-Verlag.

\bibitem[Eppstein et~al.(2007)Eppstein, Falmagne, and S.]{eppst07b}
D.~Eppstein, J.-Cl. Falmagne, and Ovchinnikov S.
\newblock \emph{{Media Theory}}.
\newblock Springer-Verlag, Berlin, Heidelberg, and New York, 2007.

\bibitem[Falmagne and Doignon(1988)]{falmagne:88a}
J.-Cl. Falmagne and J-P. Doignon.
\newblock {A class of stochastic procedures for the assessment of knowledge}.
\newblock \emph{British Journal of Mathematical and Statistical Psychology},
  41:\penalty0 1--23, 1988.

\bibitem[Falmagne and Ovchinnikov(2002)]{falma02}
J.-Cl. Falmagne and S.~Ovchinnikov.
\newblock {Media theory}.
\newblock \emph{Discrete Applied Mathematics}, 121:\penalty0 83--101, 2002.

\bibitem[Falmagne et~al.(1990)Falmagne, Koppen, Villano, Doignon, and
  Johanessen]{falma90}
J.-Cl. Falmagne, M.~Koppen, M.~Villano, J.-P. Doignon, and L.. Johanessen.
\newblock {Introduction to knowledge spaces: how to build, test and search
  them}.
\newblock \emph{Psychological Review}, 97:\penalty0 204--224, 1990.

\bibitem[Fredman and Khachiyan(1996)]{FreKha-Algs-96}
Michael~L. Fredman and Leonid Khachiyan.
\newblock {On the complexity of dualization of monotone disjunctive normal
  forms}.
\newblock \emph{Journal of Algorithms}, 21\penalty0 (3):\penalty0 618--628,
  1996.

\bibitem[Hopcroft and Karp(1973)]{HopKar-SJC-73}
J.E. Hopcroft and R.M. Karp.
\newblock {An $O(n^{5/2})$ algorithm for maximum matchings in bipartite
  graphs}.
\newblock \emph{SIAM J. on Computing}, 2\penalty0 (4):\penalty0 225--231, 1973.

\bibitem[Jerrum et~al.(2001)Jerrum, Sinclair, and Vigoda]{JerSinVig-STOC-01}
M.~Jerrum, A.~Sinclair, and E.~Vigoda.
\newblock {A polynomial-time approximation algorithm for the permanent of a
  matrix with non-negative entries}.
\newblock In \emph{Proc. 33rd ACM Symp. on Theory of Computing}, pages
  712--271, 2001.

\bibitem[Kempner and Levit(2003)]{KemLev-arXiv-03}
Y.~Kempner and V.E. Levit.
\newblock {Correspondence between two antimatroid algorithmic
  characterizations}.
\newblock Electronic preprint math.CO/0307013, arXiv.org, 2003.

\bibitem[Korte et~al.(1991)Korte, Lov{\'a}sz, and Schrader]{KorLovSch-91}
B.~Korte, L.~Lov{\'a}sz, and R.~Schrader.
\newblock \emph{{Greedoids}}.
\newblock Number~4 in Algorithms and Combinatorics. Springer-Verlag, 1991.

\bibitem[Thi{\'e}ry(2001)]{Thi-01}
N.~Thi{\'e}ry.
\newblock \emph{{Dynamically Adapting Knowledge Spaces}}.
\newblock PhD thesis, Univ. of California, Irvine, School of Social Sciences,
  2001.

\bibitem[Yannakakis(1982)]{Yan-SIADM-82}
M.~Yannakakis.
\newblock {The complexity of the partial order dimension problem}.
\newblock \emph{SIAM J. Alg. Disc. Meth.}, 3\penalty0 (3):\penalty0 351--358,
  September 1982.

\end{thebibliography}

\end{document}